# A group-theoretic approach to the bifurcation analysis of elastic frameworks with symmetry

Christelle J. Combescure[a,b,*], Timothy J. Healey[c,d], Jay Treacy[e]

[a] CREC Saint-Cyr, Acad  emie Militaire de Saint-Cyr Coetquidan, F-56380, Guer, France.
[b] IRDL, Univ. Bretagne Sud, UMR CNRS 6027, F-56100, Lorient, France.
christelle.combescure@univ-eiffel.fr
[c] Department of Mathematics, Cornell University, Ithaca, NY 15850, USA
[d] Field of Theoretical & Applied Mechanics, Cornell University
tjh10@cornell.edu
[e] jatreacy@gmail.com

**Abstract.** We present a general approach to the bifurcation analysis of elastic frameworks with symmetry. While group-theoretic methods for bifurcation problems with symmetry are well known, their actual implementation in the context of elastic frameworks is not straightforward. We consider frames comprising assemblages of Cosserat rods, and the main difficulty arises from the nonlinear configuration space, due to the presence of (cross-sectional) rotation fields. We avoid this via a single-rod formulation, developed earlier by one of the authors, whereby the governing equations are embedded in a linear space. The field equations comprise the assembly of all rod equations, supplemented by compatibility and equilibrium conditions at the joints. We demonstrate their equivariance under the symmetry-group action, and the implementation of group-theoretic methods is now natural within the linear-space context. All potential generic, symmetry-breaking bifurcations are predicted a-priori. We then employ an open-source path-following code, which can detect and compute simple, one-dimensional bifurcations; multiple bifurcation points are beyond its capabilities. For the latter, we construct symmetry-reduced problems implemented by appropriate substructures. Multiple bifurcations are rendered simple, and the path-following code is again applicable. We first analyze a simple tripod framework, providing all details of our methodology. We then treat a hexagonal space frame via the same approach. The tripod and the hexagonal dome both exhibit simple and double bifurcation points.

## Introduction

Group-theoretic methods in bifurcation problems with symmetry are well established. Sattinger was the first to introduce such ideas in the context of local bifurcation analysis [19]. A more systematic and generic-classification approach was subsequently developed by Golubitsky and co-workers [9]. Other monographs on the subject have also appeared, e.g., [2], [15], [21]. The use of group-theoretic ideas in global-numerical path following was first introduced in [10], with several other works of this type following subsequently, e.g., [5], [14]. The analyses of either simplified space-truss structures or “n-box” reaction-diffusion problems are presented in these papers. We mention that far less trivial examples of group-theoretic methods in bifurcation problems from solid/structural mechanics are presented in [3], [4], [7], [22].

We treat another class of nontrivial problems from nonlinear structural mechanics in this work, viz., elastic frameworks comprising assemblages of special Cosserat rods. The main difficulty here is that the description of even a single rod requires a nonlinear configuration-space field: The orientation of each cross-section entails a rotation, i.e., an element of SO(3). A finite element approach to the statics of an elastic Cosserat rod is presented in [20]; Newton's method is carried out on the discretization of the nonlinear solution manifold. A simple alternative approach is provided in [13]: Quaternions are directly employed in lieu of rotations, while the mixed formulation incorporates a "dummy" parameter such that the unit-quaternion equation need only be enforced at the boundary. In this way, the problem is embedded in a linear-space framework. The number of boundary conditions (seven at each end) match the correct number of unknowns (fourteen) in well-formulated rod problems. Any standard 2-point-boundary-value solver can be employed as a consequence.

[*]Corresponding author

The formulation in [13] lends itself naturally to the analysis of elastic frameworks. This has already been observed and implemented in [23]. In particular, it functions well with the open-source continuation code AUTO [5], also employed in [22] without symmetry considerations. In this work we demonstrate and exploit the equivariance of the governing equilibrium equations via group-theoretic ideas. AUTO can detect and compute simple, one-dimensional bifurcations, while multiple bifurcations points are beyond its capabilities; group-theoretic reduction becomes important. The main idea is to systematically obtain reduced problems such that multiple bifurcations from the full problem become simple.

The outline of the work is as follows. In Section 1 we first recall the equilibrium field equations for hyperelastic Cosserat rods, including their derivation via stationary potential energy. The latter is non-trivial, due to the presence of rotations; although ostensibly well known, we derive it here for completeness. More importantly, we use it to deduce group equivariance later in the work. We finish this section with a summary of the formulation presented in [13].

In Section 2 we consider a simple three-rod tripod, arranged symmetrically with a rigid or "welded" joint at the crest, where a dead load along the axis of symmetry is applied. We analyze this simple model in order to set the stage for the more realistic and complicated framework treated in the following section. We explicitly demonstrate the invariance of the total potential energy leading to the equivariance of the field equations and illustrate the use of group-theoretic ideas in a simple setting. We first make qualitative-generic predictions concerning potential bifurcations, cf. [9], [15]. Next, we analyze the complete structure via AUTO, revealing several simple bifurcations. We then construct a reduced problem to capture a double bifurcation. The latter is facilitated via an appropriate substructure. All bifurcations found are in consonance with the qualitative predictions – with one exception: A simple, symmetry-preserving bifurcation is also uncovered. While this does not follow from generic predictions, it is a consequence of the local, simultaneous buckling of the rods, which also occurs in pin-jointed tripods [1].

In Section 3 we take up the analysis of a hexagonal framework dome. This has the appearance of the simple, finite-dimensional space-truss dome considered in [10]. But here the structural members are deformable Cosserat rods connected by rigid joints. The representation of symmetry and equivariance is far more complicated now, but we rely on the same approach employed in the analysis of the tripod. Several group-theoretic predictions for potential bifurcations are again provided. In particular, two types of double bifurcations are now possible. An analysis of the full structure via AUTO reveals several simple bifurcations in accordance with the qualitative predictions. We then construct two distinct reduced problems, capturing the double bifurcations. The former are implemented via distinct substructures.

We also point out that a similar hexagonal framework dome was analyzed as part of a 2004 PhD dissertation [18]. The approach there differs from ours in several ways. The finite-element formulation of [20] is employed, yielding a problem in a high-dimensional nonlinear space - especially so in the context of a spatial framework. The equivariance of the global field equations is not discussed (but is tacitly assumed). In addition, the work employs global numerical projections onto fix-point spaces, as in the low-dimensional problem treated in [10]. Of course, projections are valid in a linear vector space only. As such, they are imposed incrementally in [18], resulting in an expensive and rather cumbersome procedure. This portion of the work from [18] was never published.

## 1. Preliminaries

*Notation:* As is common practice, we let $\mathbb{R}^3$ denote both Euclidean 3-space and its Euclidean inner-product translate space. Vectors in $\mathbb{R}^3$ are denoted by bold-face, lower case symbols, e.g., $\mathbf{x}, \mathbf{v}$, and bold-face, upper case symbols, e.g., $\mathbf{A}, \mathbf{L}$, denote linear transformations of $\mathbb{R}^3$ into itself. The inner product between vectors $\mathbf{a}, \mathbf{b}$ is denoted $\mathbf{a} \cdot \mathbf{b}$; their right-handed cross product is written $\mathbf{a} \times \mathbf{b}$. The transpose and determinant of $\mathbf{A}$ are denoted $\mathbf{A}^T$ and $\det \mathbf{A}$, respectively. The set of all proper rotations is given by $SO(3) = \{\mathbf{Q} : \mathbf{Q}^{-1} = \mathbf{Q}^T, \det \mathbf{Q} = 1\}$. For any skew-symmetric transformation $\mathbf{\Omega}$, i.e., $\mathbf{\Omega}^T = -\mathbf{\Omega}$, there is a unique vector, written $\boldsymbol{\omega} = axial(\mathbf{\Omega})$, such that $\mathbf{\Omega}\mathbf{a} = \boldsymbol{\omega} \times \mathbf{a}$ for all vectors $\mathbf{a} \in \mathbb{R}^3$.

We now summarize the formulation for a single rod pertinent to our analysis of frameworks. Let $\{\mathbf{e}_1, \mathbf{e}_2, \mathbf{e}_3\}$ denote a fixed, right-handed orthonormal basis. We consider a straight, uniform rod occupying the reference configuration $\{x\mathbf{e}_1 : x \in [0, L]\}$. Let $\mathbf{r}(x)$ denote the position vector (with respect to some fixed origin) of the material point that occupies $x\mathbf{e}_1$ in the reference configuration, while $\mathbf{R}(x) \in SO(3)$ denotes the rotation of the cross-section of the undeformed rod contained in $span\{\mathbf{e}_2, \mathbf{e}_3\}$ at position $x\mathbf{e}_1$. The field $(\mathbf{r}, \mathbf{R})$ on $[0, L]$ uniquely specifies the configuration of the rod in the deformed configuration.

The last two vectors of the orthonormal field

$$\mathbf{d}_i(x) := \mathbf{R}(x)\mathbf{e}_i, \ i = 1,2,3, \tag{1.1}$$

are called directors in the special Cosserat theory [A], employed here. Differentiation of (1.1) yields

$$\dot{\mathbf{d}}_i = \dot{\mathbf{R}}\mathbf{R}^T\mathbf{d}_i, i = 1,2,3 \tag{1.2}$$

where $\mathbf{K} := \dot{\mathbf{R}}\mathbf{R}^T$ is a skew-symmetric field and $(\dot{\ }) := d(\ )/dx$. Hence, there is a unique axial vector field

$$\boldsymbol{\kappa} = axial(\mathbf{K}) \tag{1.3}$$

such that (1.2) is equivalent to $\dot{\mathbf{d}}_i = \boldsymbol{\kappa} \times \mathbf{d}_i, \ i = 1,2,3.$ We then write

$$\dot{\mathbf{r}} = \nu_i \mathbf{d}_i \text{ and } \boldsymbol{\kappa} = \kappa_i \mathbf{d}_i. \tag{1.4}$$

The six components $\nu_i, \kappa_i, i = 1,2,3,$ comprise the strain measures of the theory. We assume that the rod is *hyperelastic*, viz., there is smooth stored-energy function $W(\nu_1, \nu_2, \nu_3, \kappa_1, \kappa_2, \kappa_3)$, $W : (0, \infty) \times \mathbb{R}^5 \to \mathbb{R}$, such that

$$\mathbf{n} = \frac{\partial W}{\partial \nu_i}\mathbf{d}_i \quad \text{and} \quad \mathbf{m} = \frac{\partial W}{\partial \kappa_i}\mathbf{d}_i \tag{1.5}$$

correspond to the contact force and contact couple fields, respectively. Moreover, the total internal potential energy of the rod is given by

$$\mathcal{U}[\mathbf{r},\mathbf{R}] := \int_0^L W(\nu_1,\nu_2,\nu_3,\kappa_1,\kappa_2,\kappa_3)dx = \int_0^L W(\mathbf{R}^T\dot{\mathbf{r}},\mathbf{R}^T\boldsymbol{\kappa})dx, \tag{1.6}$$

where we have made use of (1.1) and (1.4) in obtaining the right side of (1.6). For simplicity, we assume the absence of external body and body-couple fields. The following is well-known but not well-documented in the literature:

**Proposition 1.1** *The first-variation condition (stationary potential energy) for* (1.6) *formally delivers the balance-of-forces and balance-of-moments equilibrium equations*

$$\dot{\mathbf{n}} = \mathbf{0} \ \text{ and } \ \dot{\mathbf{m}} + \dot{\mathbf{r}} \times \mathbf{n} = \mathbf{0} \text{ in } (0,L), \tag{1.7}$$

*respectively, with* $\mathbf{n}$ and $\mathbf{m}$ *given by* (1.5).

**Proof.** For simplicity, we suppose that configuration of the rod is prescribed at the two ends, viz., $\mathbf{r}(0),\mathbf{R}(0),\mathbf{r}(L)$ and $\mathbf{R}(L)$ are specified. We consider smooth but arbitrary fields $\boldsymbol{\eta}(x),\boldsymbol{\Theta}(x)$ that vanish at the end points with $\boldsymbol{\Theta}^T \equiv -\boldsymbol{\Theta}$, i.e., $\boldsymbol{\Theta}$ is a skew-symmetric field. The former is an *additive* admissible variation for the field $\mathbf{r}$, in contrast to the variation of the rotation field $\mathbf{R}$. Rather, a finite perturbation of $\mathbf{R}(s)$ reads $\exp[\varepsilon\boldsymbol{\Theta}(s)]\mathbf{R}(s)$, and an admissible first-order variation follows from $\mathbf{R} \to \frac{d}{d\varepsilon}\exp[\varepsilon\boldsymbol{\Theta}]\mathbf{R}\,|_{\varepsilon=0} = \boldsymbol{\Theta}\mathbf{R}$. The first-variation condition for (1.6) then follows from the calculation

$$\frac{d}{d\varepsilon}\mathcal{U}[\mathbf{r}+\varepsilon\boldsymbol{\eta},\mathbf{R}+\varepsilon\boldsymbol{\Theta}\mathbf{R}+o(\varepsilon)]\,|_{\varepsilon=0} = 0 \text{ for all } \boldsymbol{\eta},\boldsymbol{\Theta}. \tag{1.8}$$

From (1.6), this entails unravelling the pertinent first-order perturbations:

$$\begin{aligned}[\mathbf{R}+\varepsilon\boldsymbol{\Theta}\mathbf{R}+o(\varepsilon)]^T(\dot{\mathbf{r}}+\varepsilon\dot{\boldsymbol{\eta}}) &= \mathbf{R}^T\dot{\mathbf{r}} + \varepsilon\mathbf{R}^T(\dot{\boldsymbol{\eta}} - \boldsymbol{\Theta}\dot{\mathbf{r}}) + o(\varepsilon) \\ &= \mathbf{R}^T\dot{\mathbf{r}} + \varepsilon\mathbf{R}^T(\dot{\boldsymbol{\eta}} + \dot{\mathbf{r}}\times\boldsymbol{\theta}) + o(\varepsilon),\end{aligned} \tag{1.9}$$

where $\boldsymbol{\theta} = axial(\boldsymbol{\Theta})$. In addition,

$$\begin{aligned}&[\dot{\mathbf{R}} + \varepsilon(\dot{\boldsymbol{\Theta}}\mathbf{R} + \boldsymbol{\Theta}\dot{\mathbf{R}}) + o(\varepsilon)][\mathbf{R}+\varepsilon\boldsymbol{\Theta}\mathbf{R}+o(\varepsilon)]^T \\ &\qquad = \mathbf{K} + \varepsilon(\dot{\boldsymbol{\Theta}} + \boldsymbol{\Theta}\mathbf{K} - \mathbf{K}\boldsymbol{\Theta}) + o(\varepsilon), \\ &axial\{\mathbf{K} + \varepsilon(\dot{\boldsymbol{\Theta}} + \boldsymbol{\Theta}\mathbf{K} - \mathbf{K}\boldsymbol{\Theta}) + o(\varepsilon)\} = \boldsymbol{\kappa} + \varepsilon(\dot{\boldsymbol{\theta}} + \boldsymbol{\theta}\times\boldsymbol{\kappa}) + o(\varepsilon),\end{aligned}$$

and then

$$\begin{aligned}&[\mathbf{R}+\varepsilon\boldsymbol{\Theta}\mathbf{R}+o(\varepsilon)]^T[\boldsymbol{\kappa} + \varepsilon(\dot{\boldsymbol{\theta}} + \boldsymbol{\theta}\times\boldsymbol{\kappa}) + o(\varepsilon)] \\ &\qquad = \mathbf{R}^T\boldsymbol{\kappa} + \varepsilon\mathbf{R}^T(\dot{\boldsymbol{\theta}} + \boldsymbol{\theta}\times\boldsymbol{\kappa} - \boldsymbol{\Theta}\boldsymbol{\kappa}) + o(\varepsilon) \\ &\qquad = \mathbf{R}^T\boldsymbol{\kappa} + \varepsilon\mathbf{R}^T\dot{\boldsymbol{\theta}} + o(\varepsilon).\end{aligned} \tag{1.10}$$

Returning to (1.8), we now find

$$\begin{aligned}\langle\delta\mathcal{U}[\mathbf{r},\mathbf{R}],(\boldsymbol{\eta},\boldsymbol{\theta})\rangle &:= \frac{d}{d\varepsilon}\mathcal{U}[\mathbf{r}+\varepsilon\boldsymbol{\eta},\mathbf{R}+\varepsilon\boldsymbol{\Theta}\mathbf{R}+o(\varepsilon)]\,|_{\varepsilon=0} \\ &= \frac{d}{d\varepsilon}\int_0^L W\left(\mathbf{R}^T\dot{\mathbf{r}} + \varepsilon\mathbf{R}^T(\dot{\boldsymbol{\eta}} + \dot{\mathbf{r}}\times\boldsymbol{\theta}) + o(\varepsilon), \mathbf{R}^T\boldsymbol{\kappa} + \varepsilon\mathbf{R}^T\dot{\boldsymbol{\theta}} + o(\varepsilon)\right)dx \\ &= \int_0^L \left\{\frac{\partial W}{\partial \nu_i}(\mathbf{R}^T\dot{\mathbf{r}},\mathbf{R}^T\boldsymbol{\kappa})\mathbf{e}_i\cdot[\mathbf{R}^T(\dot{\boldsymbol{\eta}}+\dot{\mathbf{r}}\times\boldsymbol{\theta})] + \frac{\partial W}{\partial \kappa_i}(\mathbf{R}^T\dot{\mathbf{r}},\mathbf{R}^T\boldsymbol{\kappa})\mathbf{e}_i\cdot\mathbf{R}^T\dot{\boldsymbol{\theta}}\right\}dx \\ &= \int_0^L[\mathbf{n}\cdot(\dot{\boldsymbol{\eta}}+\dot{\mathbf{r}}\times\boldsymbol{\theta}) + \mathbf{m}\cdot\dot{\boldsymbol{\theta}}]ds = 0 \ \text{ for all } \boldsymbol{\eta},\boldsymbol{\theta},\end{aligned} \tag{1.11}$$

where the last line follows from the use of (1.1) and (1.5). The first-variation condition (1.11) expresses the weak form of the equilibrium equations (1.7); the classical form (1.7) follows from (1.11) via integration by parts and the usual formal arguments of the calculus of variations.

Next, we suppose that the uniform rod has cross-sectional reflection symmetry across the plane spanned by $\{\mathbf{e}_1, \mathbf{e}_3\}$. Let $\mathbf{E}$ denote the reflection transformation defined by

$$\mathbf{E}\mathbf{e}_1 = \mathbf{e}_1, \mathbf{E}\mathbf{e}_2 = -\mathbf{e}_2, \mathbf{E}\mathbf{e}_3 = \mathbf{e}_3. \tag{1.12}$$

From the formulation in [12], a hyperelastic rod with such symmetry satisfies the invariance condition

$$W(\mathbf{E}\mathbf{R}^T\mathbf{v}, -\mathbf{E}\mathbf{R}^T\boldsymbol{\kappa}) \equiv W(\mathbf{R}^T\mathbf{v}, \mathbf{R}^T\boldsymbol{\kappa}). \tag{1.13}$$

The negative sign in (1.13) follows from the fact that $-\mathbf{E}\boldsymbol{\omega}$ is the axial vector of $\mathbf{E}\boldsymbol{\Omega}\mathbf{E}$ for any skew-symmetric transformation $\boldsymbol{\Omega}$ with axial vector $\boldsymbol{\omega}$.

Finally, we summarize the formulation presented in [13], which we employ in our forthcoming analyses. First, we express (1.7) in componential form relative to the director basis field $\{\mathbf{d}_1, \mathbf{d}_2, \mathbf{d}_3\}$. We write such components as triples of the form $\underline{n} = (n_1, n_2, n_3)$, $\underline{\nu} = (\nu_1, \nu_2, \nu_3)$, etc. Then (1.7) is equivalent to

$$\begin{aligned} \dot{\underline{n}} &= \underline{n} \times \underline{\kappa}, \\ \dot{\underline{m}} &= \underline{m} \times \underline{\kappa} + \underline{n} \times \underline{\nu}. \end{aligned} \tag{1.14}$$

We assume that the $6 \times 6$ Hessian matrix $D^2W(\cdot)$ is positive definite for each of its arguments on $(0,\infty) \times \mathbb{R}^5$, which insures the existence of a complementary energy density $\Gamma(\underline{n}, \underline{m})$ (the Legendre transform of $W$) such that

$$\underline{\nu} = \frac{\partial \Gamma}{\partial \underline{n}} \text{ and } \underline{\kappa} = \frac{\partial \Gamma}{\partial \underline{m}}. \tag{1.15}$$

The substitution of (1.15) into (1.14) gives

$$\begin{aligned} \dot{\underline{n}} &= \underline{n} \times \frac{\partial \Gamma}{\partial \underline{m}}, \\ \dot{\underline{m}} &= \underline{m} \times \frac{\partial \Gamma}{\partial \underline{m}} + \underline{n} \times \frac{\partial \Gamma}{\partial \underline{n}}. \end{aligned} \tag{1.16}$$

Next, we express the placement fields $\mathbf{r}, \mathbf{R}$ in terms of their components with respect to the fixed basis, viz.,

$$\begin{aligned} \mathbf{r} &= r_i\mathbf{e}_i, \ \bar{r} := (r_1, r_2, r_3), \\ \mathbf{R} &= R_{ij}\mathbf{e}_i \otimes \mathbf{e}_j, \bar{R} := [R_{ij}], \end{aligned} \tag{1.17}$$

where $[R_{ij}]$ represents the $3 \times 3$ matrix of components. We henceforth express all components with respect to the fixed basis with an overbar, as above in (1.17). According to Euler's theorem, there is an "instantaneous" axis of rotation corresponding to a real eigenvector $\mathbf{n}$ with $\mathbf{Rn} = \mathbf{n}$. Let $\phi \in \mathbb{R}(\bmod 2\pi)$ denote the counterclockwise angle of rotation about $\mathbf{n}$ (according to the right-hand rule), given by the trace formula $tr\mathbf{R} = 1 + 2\cos\phi$. Choosing $\mathbf{n} \cdot \mathbf{n} = 1$, we associate $\mathbf{R}$ with the quaternion

$$\begin{aligned} \mathsf{q} &= (q_o, \mathbf{q}), \\ q_o &:= \cos(\phi/2), \ \mathbf{q} := \sin(\phi/2)\mathbf{n}. \end{aligned} \tag{1.18}$$

Writing $\mathbf{q} = q_i \mathbf{e}_i$, it follows that

$$q_o^2 + \mathbf{q} \cdot \mathbf{q} = q_o^2 + q_1^2 + q_2^2 + q_3^2 = 1, \tag{1.19}$$

and the matrix $\bar{R}$ admits the parametrization

$$\bar{R}(\bar{\mathrm{q}}) = 2 \begin{bmatrix} q_o^2 + q_1^2 - 1/2 & q_1 q_2 - q_o q_3 & q_1 q_3 + q_o q_2 \\ q_1 q_2 + q_o q_3 & q_o^2 + q_2^2 - 1/2 & q_2 q_3 - q_o q_1 \\ q_1 q_3 - q_o q_2 & q_2 q_3 + q_o q_1 & q_o^2 + q_3^2 - 1/2 \end{bmatrix}, \tag{1.20}$$

Where $\bar{\mathrm{q}} := (q_o, q_1, q_2, q_3)$. Finally, as derived in [13], we have

$$\begin{aligned} \dot{\bar{r}} &= \bar{R}(\bar{\mathrm{q}})\underline{\nu} = \bar{R}(\bar{\mathrm{q}})\frac{\partial \Gamma}{\partial \underline{n}}, \\ \dot{\bar{\mathrm{q}}} &= \mathrm{A}(\bar{\mathrm{q}})\underline{\kappa} + \mu \bar{\mathrm{q}} = \mathrm{A}(\bar{\mathrm{q}})\frac{\partial \Gamma}{\partial \underline{n}} + \mu \bar{\mathrm{q}}, \end{aligned} \tag{1.21}$$

where $\mu \in \mathbb{R}$ is an unspecified (dummy) parameter and

$$\mathrm{A}(\bar{\mathrm{q}}) := \begin{bmatrix} -q_1 & -q_2 & -q_3 \\ q_o & -q_3 & q_2 \\ q_3 & q_o & -q_1 \\ -q_2 & q_1 & q_o \end{bmatrix}. \tag{1.22}$$

The field equations comprise (1.16) and (1.21). As shown in [12], if (1.19) is satisfied at the endpoints, then $\mu$ is zero and (1.19) automatically holds over $(0, L)$, i.e., (1.19) is not a required field equation. For a rotational boundary condition of place, (1.19) is automatically satisfied at that boundary point. On the other hand, if an applied couple is imposed or if a "mixed" rotational boundary condition is prescribed, then (1.19) must be specified at that boundary point as well. Either way, seven boundary conditions are always prescribed at each end, which agrees with the fourteen unknowns $\underline{n}, \underline{m}, \bar{r}, \mathrm{q}$ and $\mu$. In practice via finite-precision arithmetic, $\mu$ computes to an extremely small value.

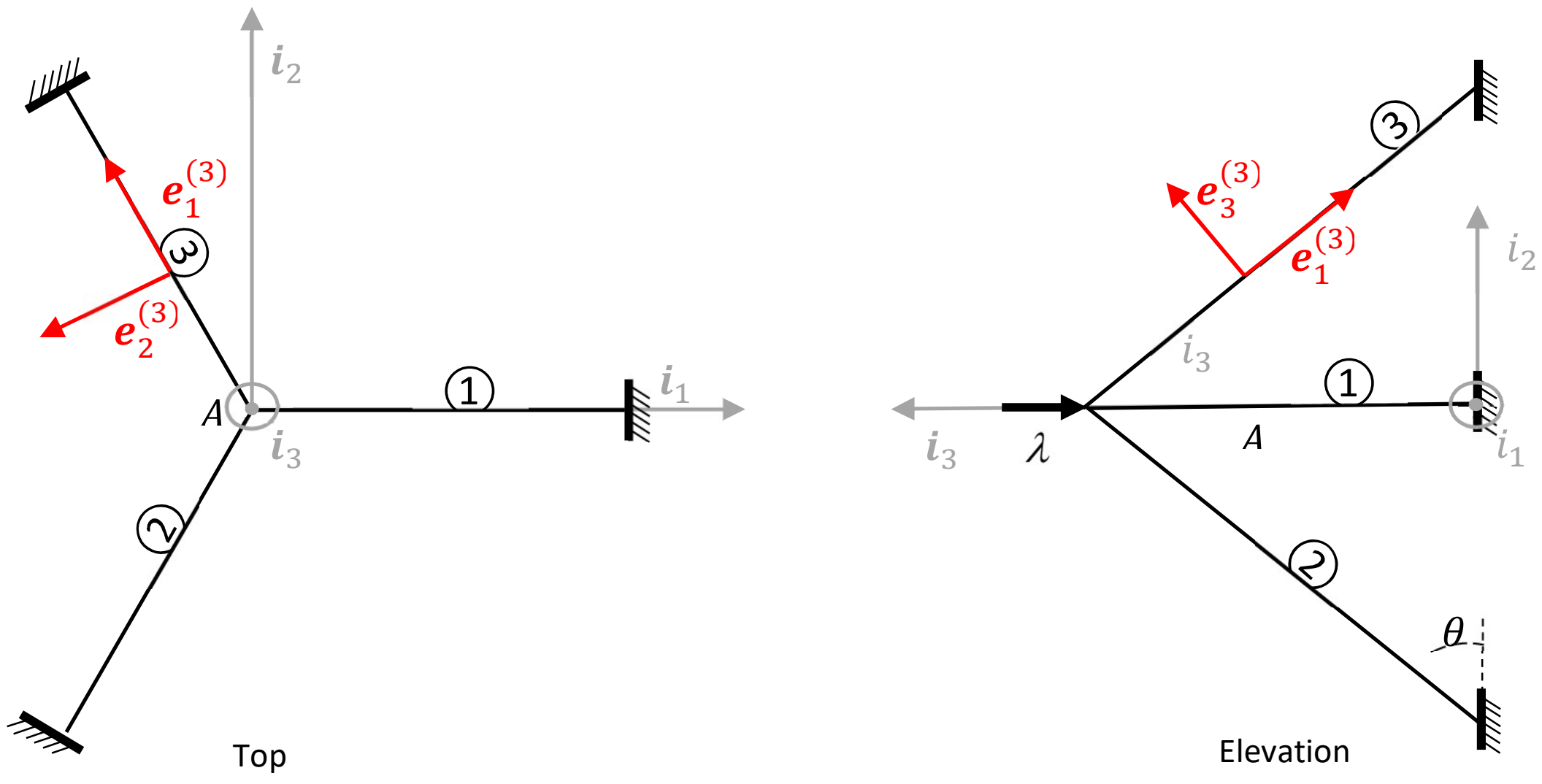

Figure 1: Global and local basis for the tripod represented on top-down (left) and elevation(right) view.

## 2. A Simple Tripod Framework with Symmetry:

### 2.1 Formulation and Equivariance

We first consider a simple tripod composed of three identical, straight, prismatic, unstressed, hyperelastic rods of length $L$ all meeting at a perfectly rigid or welded joint at A, i.e., all rods have the same displacement and rotation at the joint. The rods are arranged symmetrically, the supports of the rods are identical, and we assume that the angle between the horizontal plane and each bar is $\theta = \pi / 3$ radians, cf. Figure 1. Let $\{\mathbf{i}_1, \mathbf{i}_2, \mathbf{i}_3\}$ denote a fixed orthonormal *global* basis for $\mathbb{R}^3$ and let $\{\mathbf{e}_1^{(j)}, \mathbf{e}_2^{(j)}, \mathbf{e}_3^{(j)}\}$ be the fixed orthonormal *local* basis oriented along the $j^{th}$ undeformed rod, $j = 1,2,3,$ such that

$$\begin{gathered}\mathbf{e}_1^{(1)} = (1/2)\mathbf{i}_1 - (\sqrt{3}/2)\mathbf{i}_3,\ \mathbf{e}_2^{(1)} = \mathbf{i}_2,\ \mathbf{e}_3^{(1)} = (\sqrt{3}/2)\mathbf{i}_1 + (1/2)\mathbf{i}_3,\\ \mathbf{e}_j^{(2)} = \mathbf{R}_{2\pi/3}\mathbf{e}_j^{(1)},\ \mathbf{e}_j^{(3)} = \mathbf{R}_{2\pi/3}\mathbf{e}_j^{(2)} = \mathbf{R}_{4\pi/3}\mathbf{e}_j^{(1)},\ j = 1,2,3,\end{gathered} \tag{2.1}$$

where $\mathbf{R}_{2\pi/3}$ denotes a clockwise rotation through the angle $2\pi/3$ about $\mathbf{i}_3$; its matrix relative to the fixed global basis is given by

$$[R_{2\pi/3}] = \frac{1}{2}\begin{bmatrix} -1 & \sqrt{3} & 0 \\ -\sqrt{3} & -1 & 0 \\ 0 & 0 & 2 \end{bmatrix}. \tag{2.2}$$

In what follows, repeated subscripts sum (Einstein convention) while repeated parenthetical superscripts, labeling the rods, do not. We assume that the $j^{th}$ rod has reflection symmetry across the plane spanned by $\{\mathbf{e}_1^{(j}), \mathbf{e}_3^{(j)}\}$.

A configuration of the tripod is uniquely described by the placement and orientation field of each rod, represented by

$$\mathbf{u} := \left(\mathbf{r}^{(1)}, \mathbf{R}^{(1)}, \mathbf{r}^{(2)}, \mathbf{R}^{(2)}, \mathbf{r}^{(3)}, \mathbf{R}^{(3)}\right), \tag{2.3}$$

where $\mathbf{r}^{(j)} : [0, L] \to \mathbb{R}^3$ and $\mathbf{R}^{(j)} : [0, L] \to SO(3)$ denote the placement and rotation field of the deformed $j^{th}$ rod, respectively, $j = 1,2,3;$ each is presumed to be continuously differentiable.
For compatibility at joint A, we require

$$\begin{gathered}\mathbf{r}^{(1)}(0) = \mathbf{r}^{(2)}(0) = \mathbf{r}^{(3)}(0) := \mathbf{r}_A,\\ \mathbf{R}^{(1)}(0) = \mathbf{R}^{(2)}(0) = \mathbf{R}^{(3)}(0) := \mathbf{R}_A,\end{gathered} \tag{2.4}$$

where $\mathbf{r}_A, \mathbf{R}_A,$ the joint placement and orientation, respectively, are unknowns.

If the supported ends of the rods are clamped, we then we require

$$\begin{gathered}\mathbf{r}^{(1)}(L) = (L/2)\mathbf{i}_1,\ \mathbf{r}^{(2)}(L) = (L/2)\mathbf{R}_{2\pi/3}\mathbf{i}_1,\ \mathbf{r}^{(3)}(L) = (L/2)\mathbf{R}_{4\pi/3}\mathbf{i}_1,\\ \mathbf{R}^{(1)}(L) = \mathbf{R}^{(2)}(L) = \mathbf{R}^{(3)}(L) = \mathbf{I},\end{gathered} \tag{2.5}$$

where $\mathbf{I}$ denotes the identity. For "ball-in-socket" supports, $(2.5)_1$ still hold along with the vanishing of all end moments, viz.,

$$\mathbf{m}^{(j)}(L) = \mathbf{0}, j = 1,2,3, \tag{2.6}$$

where $\mathbf{m}^{(j)}:(0,L)\to\mathbb{R}^3$ denotes the contact-couple field for the $j^{th}$ rod, $j=1,2,3.$ In this case, an additional boundary condition, corresponding to (1.19) evaluated at $x=L,$ must also be incorporated. Of course, other compatibility/boundary conditions are possible, e.g., fixed supports (2.5) with a ball-in-socket at joint A, implying vanishing torques as in (2.6) but evaluated at $x=0.$ Other "mixed" conditions (symmetrically arranged) are possible, the explicit details of which we do not present.

We assume that the tripod is subjected to a dead, concentrated load at joint A of the form $-\lambda\mathbf{i}_3$. The total potential energy of the structure is then given by

$$\mathcal{V}[\lambda,\mathbf{u}]=\mathcal{U}[\mathbf{u}]+\lambda\mathbf{r}_A\cdot\mathbf{i}_3, \tag{2.7}$$

where

$$\mathcal{U}[\mathbf{u}]:=\sum_{j=1}^{3}\int_0^L W([\mathbf{R}^{(j)}]^T\dot{\mathbf{r}}^{(j)},[\mathbf{R}^{(j)}]^T\boldsymbol{\kappa}^{(j)})dx,$$
$$[\mathbf{R}^{(j)}]^T\dot{\mathbf{r}}^{(j)}=\nu_i^{(j)}\mathbf{e}_i^{(j)},\ [\mathbf{R}^{(j)}]^T\boldsymbol{\kappa}^{(j)}=\kappa_i^{(j)}\mathbf{e}_i^{(j)},$$

and $W:(0,\infty)\times\mathbb{R}^5\to\mathbb{R}$ is the stored-energy function, common to each of the rods. We now compute the first-variation condition. For simplicity, we assume the end conditions (2.4), (2.5). In consonance with (2.4), we consider smooth fields $\boldsymbol{\eta}^{(j)}(s),\ \boldsymbol{\Theta}^{(j)}(s)$ satisfying

$$\begin{aligned}&\boldsymbol{\eta}^{(j)}(L)=\mathbf{0},\ \boldsymbol{\Theta}^{(j)}(L)=\mathbf{O}, j=1,2,3,\\&\boldsymbol{\eta}_A:=\boldsymbol{\eta}^{(1)}(0)=\boldsymbol{\eta}^{(2)}(0)=\boldsymbol{\eta}^{(3)}(0),\\&\boldsymbol{\Theta}^{(1)}(0)=\boldsymbol{\Theta}^{(2)}(0)=\boldsymbol{\Theta}^{(3)}(0),\end{aligned} \tag{2.8}$$

which we use to construct admissible variations of $(\mathbf{r}^{(j)},\mathbf{R}^{(j)}), j=1,2,3,$ respectively as in (1.10). We express this abstractly via

$$\begin{aligned}\tilde{\mathcal{U}}[\mathbf{u};\mathbf{h},\varepsilon]:=\mathcal{U}(&\mathbf{r}^{(1)}+\varepsilon\boldsymbol{\eta}^{(1)},\mathbf{R}^{(1)}+\varepsilon\boldsymbol{\Theta}^{(1)}\mathbf{R}^{(1)},\mathbf{r}^{(2)}+\varepsilon\boldsymbol{\eta}^{(2)},\mathbf{R}^{(2)}+\varepsilon\boldsymbol{\Theta}^{(2)}\mathbf{R}^{(2)},\mathbf{r}^{(3)}+\varepsilon\boldsymbol{\eta}^{(3)},\\&\mathbf{R}^{(3)}+\varepsilon\boldsymbol{\Theta}^{(3)}\mathbf{R}^{(3)})+o(\varepsilon),\end{aligned} \tag{2.9}$$

where $\mathbf{h}:=(\boldsymbol{\eta}^{(1)},\boldsymbol{\theta}^{(1)},\boldsymbol{\eta}^{(2)},\boldsymbol{\theta}^{(2)},\boldsymbol{\eta}^{(3)},\boldsymbol{\theta}^{(3)})$, with $\boldsymbol{\theta}^{(i)}=axial(\boldsymbol{\Theta}^{(i)})$.

Then as in (1.11), we compute the first-variation condition:

$$\begin{aligned}\langle\delta\mathcal{V}[\lambda,\mathbf{u}],\mathbf{h}\rangle&:=\frac{d}{d\varepsilon}\left\{\tilde{\mathcal{U}}[\mathbf{u};\mathbf{h},\varepsilon)]+\lambda(\mathbf{r}_A+\varepsilon\boldsymbol{\eta}_A)\cdot\mathbf{i}_3\right\}\Big|_{\varepsilon=0}=\langle\delta\mathcal{U}[\lambda,\mathbf{u}],\mathbf{h}\rangle+\lambda\mathbf{i}_3\cdot\boldsymbol{\eta}_A\\&=\sum_{j=1}^{3}\Big\{\frac{d}{d\varepsilon}\int_0^L W\big([\mathbf{R}^{(j)}]^T\dot{\mathbf{r}}^{(j)}+\varepsilon[\mathbf{R}^{(j)}]^T(\dot{\boldsymbol{\eta}}^{(j)}+\dot{\mathbf{r}}^{(j)}\times\boldsymbol{\theta}^{(j)})+o(\varepsilon),\ [\mathbf{R}^{(j)}]^T\boldsymbol{\kappa}^{(j)}+\varepsilon[\mathbf{R}^{(j)}]^T\dot{\boldsymbol{\theta}}^{(j)}+o(\varepsilon)\big)dx\Big\}\ \Big|_{\varepsilon=0}\\&\qquad+\lambda\mathbf{i}_3\cdot\boldsymbol{\eta}_A\\&=\sum_{j=1}^{3}\int_0^L\Big\{\frac{\partial W}{\partial\nu_i}([\mathbf{R}^{(j)}]^T\dot{\mathbf{r}}^{(j)},[\mathbf{R}^{(j)}]^T\boldsymbol{\kappa}^{(j)})\mathbf{e}_i^{(j)}\cdot[\mathbf{R}^{(j)}]^T(\dot{\boldsymbol{\eta}}^{(j)}+\dot{\mathbf{r}}^{(j)}\times\boldsymbol{\theta}^{(j)})\\&\qquad+\frac{\partial W}{\partial\kappa_i}([\mathbf{R}^{(j)}]^T\dot{\mathbf{r}}^{(j)},[\mathbf{R}^{(j)}]^T\boldsymbol{\kappa}^{(j)})\mathbf{e}_i^{(j)}\cdot[\mathbf{R}^{(j)}]^T\dot{\boldsymbol{\theta}}^{(j)}\Big\}dx+\lambda\mathbf{i}_3\cdot\boldsymbol{\eta}_A=0\qquad(2.10)\\&=\sum_{j=1}^{3}\int_0^L\left\{\mathbf{n}^{(j)}\cdot(\dot{\boldsymbol{\eta}}^{(j)}+\dot{\mathbf{r}}^{(j)}\times\boldsymbol{\theta}^{(j)})+\mathbf{m}^{(j)}\cdot\dot{\boldsymbol{\theta}}^{(j)}\right\}dx+\lambda\mathbf{i}_3\cdot\boldsymbol{\eta}_A=0,\end{aligned}$$

for all admissible fields $\boldsymbol{\eta}^{(j)}, \boldsymbol{\theta}^{(j)} = axial\boldsymbol{\Theta}^{(j)}, j = 1,2,3,$ satisfying (2.8) ; we used (1.1) and (1.4) in going from $(2.10)_2$ to $(2.10)_3$. The last line of (2.10) expresses the weak form of the equilibrium equations. Integration by parts yields, use of (2.8), and the usual formal arguments of the calculus of variations then deliver

$$\begin{gathered}\dot{\mathbf{n}}^{(j)} = \mathbf{0} \text{ and } \dot{\mathbf{m}}^{(j)} + \dot{\mathbf{r}}^{(j)} \times \mathbf{n}^{(j)} = \mathbf{0} \text{ in } (0,L), \text{ j=1,2,3}, \\ \sum_{j=1}^{3} \mathbf{n}^{(j)}(0) - \lambda \mathbf{i}_3 = \mathbf{0}, \ \sum_{j=1}^{3} \mathbf{m}^{(j)}(0) = \mathbf{0},\end{gathered} \tag{2.11}$$

where $$\mathbf{n}^{(j)} = \frac{\partial W}{\partial \nu_i^{(j)}} \mathbf{d}_i^{(j)} \text{ and } \mathbf{m}^{(j)} = \frac{\partial W}{\partial \kappa_i^{(j)}} \mathbf{d}_i^{(j)}, j = 1,2,3.$$

In order to employ the formulation (1.16), (1.21), we first replace $(2.11)_3$ with

$$\underline{\nu}^{(j)} = \frac{\partial \Gamma}{\partial \underline{n}^{(j)}} \text{ and } \underline{\kappa}^{(j)} = \frac{\partial \Gamma}{\partial \underline{m}^{(j)}}, \ j = 1,2,3. \tag{2.12}$$

Likewise, we write the system (1.16)-(1.22) for each of the rods, denoted by a parenthetical superscript $j = 1,2,3.$ Finally, we replace (2.3), $(2.4)_2$ and $(2.5)_2$ by

$$\begin{aligned}&\mathsf{u} := \left(\mathbf{r}^{(1)}, \mathsf{q}^{(1)}, \mathbf{r}^{(2)}, \mathsf{q}^{(2)}, \mathbf{r}^{(3)}, \mathsf{q}^{(3)}\right), \\ &\mathsf{q}^{(1)}(0) = \mathsf{q}^{(2)}(0) = \mathsf{q}^{(3)}(0), \\ &\mathsf{q}^{(1)}(L) = \mathsf{q}^{(2)}(L) = \mathsf{q}^{(3)}(L) = (1, \mathbf{0}),\end{aligned} \tag{2.13}$$

respectively. We remark that a given a unit quaternion $\mathsf{q} = (q_o, \mathbf{q}),\ (q_o)^2 + \mathbf{q} \cdot \mathbf{q} = 1,$ corresponds to a unique rotation. However, there are two choices for a unit quaternion for a given rotation $\mathbf{R} \in SO(3)$: If $\mathsf{q}$ corresponds to $\mathbf{R},$ then so too does $-\mathsf{q}$. Nonetheless, this causes us no problems: Once a choice has been made, we employ only continuous unit-quaternion fields as functions of continuation parameters.

The tripod possesses the symmetry group $C_{3v}$, generated by the rotation $\mathbf{R}_{2\pi/3}$ and the reflection across the plane spanned by $\{\mathbf{i}_1, \mathbf{i}_3\}$. We denote the latter transformation by $\mathbf{E}$, which has the matrix

$$[E] = \begin{bmatrix} 1 & 0 & 0 \\ 0 & -1 & 0 \\ 0 & 0 & 1 \end{bmatrix} \tag{2.14}$$

relative to the global basis. More specifically,

$$C_{3v} = \{\mathbf{R}_{2k\pi/3}, \mathbf{E}\mathbf{R}_{2k\pi/3} : k = 1,2,3\}. \tag{2.15}$$

Observe that the transformation $\mathbf{E}\mathbf{R}_{2k\pi/3}$ represents a reflection across the plane spanned by $\{\mathbf{R}_{k\pi/3}\mathbf{i}_1, \mathbf{i}_3\}, k = 1,2,3,$ where $\mathbf{R}_{k\pi/3}$ denotes the clockwise rotation about $\mathbf{i}_3$ through the angle $k\pi/3$.

The natural group action of $C_{3v}$ on configuration space $(2.13)_1$ is generated by

$$\boldsymbol{T}_{R_{2\pi/3}}\mathsf{u} := (\mathbf{R}_{2\pi/3}\mathbf{r}^{(3)}, \mathsf{R}_{2\pi/3}\mathsf{q}^{(3)}, \mathbf{R}_{2\pi/3}\mathbf{r}^{(1)}, \mathsf{R}_{2\pi/3}\mathsf{q}^{(1)}, \mathbf{R}_{2\pi/3}\mathbf{r}^{(2)}, \mathsf{R}_{2\pi/3}\mathsf{q}^{(2)})$$

and (2.16)

$$\boldsymbol{T}_E\mathsf{u} := (\mathbf{E}\mathbf{r}^{(1)}, \mathsf{E}\mathsf{q}^{(1)}, \mathbf{E}\mathbf{r}^{(3)}, \mathsf{E}\mathsf{q}^{(3)}, \mathbf{E}\mathbf{r}^{(2)}, \mathsf{E}\mathsf{q}^{(2)}),$$

where the actions on unit quaternions are defined as follows:

$$\begin{aligned} \mathsf{R}_{2\pi/3}\mathsf{q} &:= (q_o, \mathbf{R}_{2\pi/3}\mathbf{q}), \\ \mathsf{E}\mathsf{q} &:= (q_o, -\mathbf{E}\mathbf{q}). \end{aligned} \tag{2.17}$$

The group generated by the operations (2.16) is a faithful representation of $C_{3v}$ on configuration space. Observe that $(2.16)_1$ corresponds to a rotation of a configuration (2.3) about $\mathbf{i}_3$ through the clockwise angle $2\pi/3$, while $(2.16)_2$ represents a reflection of a configuration across the plane spanned by $\{\mathbf{i}_1, \mathbf{i}_3\}$. The other elements in the group $\{\boldsymbol{T}_{R_{2j\pi/3}}, \boldsymbol{T}_E\boldsymbol{T}_{R_{2j\pi/3}} = \boldsymbol{T}_{ER_{2j\pi/3}}, j = 1,2,3\}$ have similar interpretations. For that reason, the representation is termed *natural*.

Our first task here is to establish:

**Proposition 2.1** *The potential energy* (2.7) *is invariant under the representation generated by the transformations* (2.16), i.e.,

$$\mathcal{V}[\lambda, \boldsymbol{T}_G\mathsf{u}] = \mathcal{V}[\lambda, \mathsf{u}] \text{ for all } \mathbf{G} \in C_{3v}. \tag{2.18}$$

**Proof:** It is enough to show that (2.18) holds for the generators (2.16). First notice that $(2.16)_1$ involves

$$\mathbf{r}^{(j)} \to \mathbf{R}_{2\pi/3}\mathbf{r}^{(j)} \text{ and } \mathsf{q}^{(j)} \to \mathsf{R}_{2\pi/3}\mathsf{q}^{(j)} \Rightarrow \mathbf{R}^{(j)} \to \mathbf{R}_{2\pi/3}\mathbf{R}^{(j)}\mathbf{R}^T_{2\pi/3}, \; j = 1,2,3. \tag{2.19}$$

In view of (2.1) and $(2.7)_2$, we deduce from (2.19) that

$$[\mathbf{R}^{(j)}]^T\mathbf{r}^{(j)} \to \nu_i^{(j)}\mathbf{e}_i^{(j+1)}, j = 1,2,3, \tag{2.20}$$

with the understanding that $\mathbf{e}_i^{(4)} \equiv \mathbf{e}_i^{(1)}, i = 1,2,3.$ Obviously this goes the same way for the last three arguments of $W$, viz.,

$$\boldsymbol{\kappa}^{(j)} \to \mathbf{R}_{2\pi/3}\boldsymbol{\kappa}^{(j)} \text{ and } \mathbf{R}^{(j)} \to \mathbf{R}_{2\pi/3}\mathbf{R}^{(j)}\mathbf{R}^T_{2\pi/3} \Rightarrow [\mathbf{R}^{(j)}]^T\boldsymbol{\kappa}^{(j)} \to \kappa_i^{(j)}\mathbf{e}_i^{(j+1)}, j = 1,2,3. \tag{2.21}$$

Then (2.7), $(2.16)_1$, (2.20) and (2.21) imply

$$\mathcal{V}[\lambda, \boldsymbol{T}_{R_{2\pi/3}}\mathsf{u}] = \sum_{j=1}^{3}\int_0^L W(\nu_i^{(j)}\mathbf{e}_i^{(j+1)}, \; \kappa_i^{(j)}\mathbf{e}_i^{(j+1)})dx + \lambda\mathbf{R}_{2\pi/3}\mathbf{r}_A \cdot \mathbf{i}_3. \tag{2.22}$$

In particular, the strains from the $j^{th}$ rod "move" to the $(j+1)^{th}$ rod under the transformation, and the invariance of the internal potential energy is now clear. The loading term is also invariant: $\mathbf{R}_{2\pi/3}\mathbf{i}_3 = \mathbf{i}_3$ and $\mathbf{R}_{2\pi/3}\mathbf{r}_A \cdot \mathbf{R}_{2\pi/3}\mathbf{i}_3 = \mathbf{r}_A \cdot \mathbf{i}_3$. Given that the compatibility and boundary conditions (2.4), (2.5), and (2.13) are unaffected by $(2.16)_1$, we conclude that (2.18) holds for $\mathbf{G} = \mathbf{R}_{2\pi/3}$.

The demonstration for $\boldsymbol{T}_E$ requires more care, due to the reflection symmetry of each individual rod. The first step is similar to (2.19):

$$\mathbf{r}^{(j)} \to \mathbf{E}\mathbf{r}^{(j)} \text{ and } \mathsf{q}^{(j)} \to \mathsf{E}\mathsf{q}^{(j)} \Rightarrow \mathbf{R}^{(j)} \to \mathbf{E}\mathbf{R}^{(j)}\mathbf{E}, \; j = 1,2,3. \tag{2.23}$$

Hence,

$$\begin{aligned} &[\mathbf{R}^{(j)}]^T\dot{\mathbf{r}}^{(j)} \to \mathbf{E}[\mathbf{R}^{(j)}]^T\dot{\mathbf{r}}^{(j)}, \\ &[\mathbf{R}^{(j)}]^T\boldsymbol{\kappa}^{(j)} = axial([\mathbf{R}^{(j)}]^T\dot{\mathbf{R}}^{(j)}) \to axial(\mathbf{E}[\mathbf{R}^{(j)}]^T\dot{\mathbf{R}}^{(j)}\mathbf{E}) = -\mathbf{E}[\mathbf{R}^{(j)}]^T\boldsymbol{\kappa}^{(j)}. \end{aligned} \tag{2.24}$$

Next, we note

$$\begin{aligned}&\mathbf{E}\mathbf{e}_1^{(1)}=\mathbf{e}_1^{(1)},\mathbf{E}\mathbf{e}_2^{(1)}=-\mathbf{e}_2^{(1)},\mathbf{E}\mathbf{e}_3^{(1)}=\mathbf{e}_3^{(1)},\\&\mathbf{E}\mathbf{e}_1^{(2)}=\mathbf{e}_1^{(3)},\mathbf{E}\mathbf{e}_2^{(2)}=-\mathbf{e}_2^{(3)},\mathbf{E}\mathbf{e}_3^{(2)}=\mathbf{e}_3^{(3)}.\end{aligned}\tag{2.25}$$

Then (2.7), $(2.16)_2$, (2.23)-(2.25) yield

$$\begin{aligned}\mathcal{V}[\lambda,\boldsymbol{T}_E\mathbf{u}]=&\int_0^L W(\mathbf{E}^{(1)}[\nu_i^{(1)}\mathbf{e}_i^{(1)}],-\mathbf{E}^{(1)}[\kappa_i^{(1)}\mathbf{e}_i^{(1)}])dx\\&+\int_0^L W(\mathbf{E}^{(2)}[\nu_1^{(3)}\mathbf{e}_1^{(2)}+\nu_2^{(3)}\mathbf{e}_2^{(2)}+\nu_3^{(3)}\mathbf{e}_3^{(2)}],-\mathbf{E}^{(2)}[\kappa_1^{(3)}\mathbf{e}_1^{(2)}+\kappa_2^{(3)}\mathbf{e}_2^{(2)}+\kappa_3^{(3)}\mathbf{e}_3^{(2)}])dx\\&+\int_0^L W(\mathbf{E}^{(3)}[\nu_1^{(2)}\mathbf{e}_1^{(3)}+\nu_2^{(2)}\mathbf{e}_2^{(3)}+\nu_3^{(2)}\mathbf{e}_3^{(3)}],-\mathbf{E}^{(3)}[\kappa_1^{(2)}\mathbf{e}_1^{(3)}+\kappa_2^{(2)}\mathbf{e}_2^{(3)}+\kappa_3^{(2)}\mathbf{e}_3^{(3)}])dx\\&+\lambda\mathbf{E}\mathbf{r}_A\cdot\mathbf{i}_3,\end{aligned}\tag{2.26}$$

where $\mathbf{E}^{(j)}$ denotes the reflection across the plane spanned by $\{\mathbf{e}_1^{(j)},\mathbf{e}_3^{(j)}\}, j=1,2,3.$ Now use the fact that $\mathbf{E}\mathbf{i}_3=\mathbf{i}_3$ along with cross-sectional reflection symmetry (1.13) to see that (2.26) becomes

$$\begin{aligned}\mathcal{V}[\lambda,\boldsymbol{T}_E\mathbf{u}]=&\int_0^L W(\nu_i^{(1)}\mathbf{e}_i^{(1)},\ \kappa_i^{(1)}\mathbf{e}_i^{(1)})dx\\&+\int_0^L W(\nu_1^{(3)}\mathbf{e}_1^{(2)}+\nu_2^{(3)}\mathbf{e}_2^{(2)}+\nu_3^{(3)}\mathbf{e}_3^{(2)},\ \kappa_1^{(3)}\mathbf{e}_1^{(2)}+\kappa_2^{(3)}\mathbf{e}_2^{(2)}+\kappa_3^{(3)}\mathbf{e}_3^{(2)})dx\\&+\int_0^L W(\nu_1^{(2)}\mathbf{e}_1^{(3)}+\nu_2^{(2)}\mathbf{e}_2^{(3)}+\nu_3^{(2)}\mathbf{e}_3^{(3)},\ \kappa_1^{(2)}\mathbf{e}_1^{(3)}+\kappa_2^{(2)}\mathbf{e}_2^{(3)}+\kappa_3^{(2)}\mathbf{e}_3^{(3)}])dx\\&+\lambda\mathbf{r}_A\cdot\mathbf{i}_3,\end{aligned}\tag{2.27}$$

In this case, the strains from rod 3 move to rod 2 and vice-versa under the transformation. It is straightforward to demonstrate that the compatibility and boundary conditions (2.4), (2.5) and (2.13) are unaffected by $(2.16)_2$. We conclude that (2.27) is precisely the same as $\mathcal{V}[\lambda,\mathbf{u}]$. □

We now use the invariance of the potential energy (2.18) to establish the *equivariance* of the governing equations (2.10), the proof of which is provided in Appendix A.

**Proposition 2.2** *The equilibrium equations (2.10) are equivariant, i.e.,*

$$\left\langle\delta\mathcal{V}[\lambda,\mathbf{T}_G\mathbf{u}],\mathbf{h}\right\rangle=\left\langle\tilde{\mathbf{T}}_G\delta\mathcal{V}[\lambda,\mathbf{u}],\mathbf{h}\right\rangle,\tag{2.28}$$

*for all* $\mathbf{G}\in C_{3v}$ *and for all admissible* $\mathbf{h}$, *as defined in* (2.9). *Here the generators of the action of* $\tilde{\boldsymbol{T}}_G$ *on* $\boldsymbol{\sigma}:=\delta\mathcal{V}$ *are defined by*

$$\begin{aligned}&\tilde{\boldsymbol{T}}_{R_{2\pi/3}}\boldsymbol{\sigma}:=(\mathbf{R}_{2\pi/3}\mathbf{n}^{(3)},\mathbf{R}_{2\pi/3}\mathbf{m}^{(3)},\mathbf{R}_{2\pi/3}\mathbf{n}^{(1)},\mathbf{R}_{2\pi/3}\mathbf{m}^{(1)},\mathbf{R}_{2\pi/3}\mathbf{n}^{(2)},\mathbf{R}_{2\pi/3}\mathbf{m}^{(2)}),\\&\tilde{\boldsymbol{T}}_E\boldsymbol{\sigma}:=(\mathbf{E}\mathbf{n}^{(1)},-\mathbf{E}\mathbf{m}^{(1)},\mathbf{E}\mathbf{n}^{(3)},-\mathbf{E}\mathbf{m}^{(3)},\mathbf{E}\mathbf{n}^{(2)},-\mathbf{E}\mathbf{m}^{(2)}),\end{aligned}\tag{2.29}$$

*for all* $\boldsymbol{\sigma}=(\mathbf{n}^{(1)},\mathbf{m}^{(1)},\mathbf{n}^{(2)},\mathbf{m}^{(2)},\mathbf{n}^{(3)},\mathbf{m}^{(3)})$, *and the inner product employed in (2.28) is defined by*

$$\left\langle\boldsymbol{\sigma},\mathbf{h}\right\rangle:=\sum_{j=1}^{3}\int_0^L\left\{\mathbf{n}^{(j)}\cdot\boldsymbol{\eta}^{(j)}+\mathbf{m}^{(j)}\cdot\boldsymbol{\theta}^{(j)}\right\}dx.\tag{2.30}$$

We express (2.28) abstractly in the strong form via

$$\mathbf{f}(\lambda,\boldsymbol{T}_G\mathbf{u})=\tilde{\boldsymbol{T}}_G\mathbf{f}(\lambda,\mathbf{u})\text{ for all }\mathbf{G}\in C_{3v},\tag{2.31}$$

where

$$\mathbf{f}(\lambda,\mathbf{u})=\mathbf{0} \tag{2.32}$$

represents the strong form of the equilibrium equations (2.11).

## 2.2 Exploitation of Symmetry and Bifurcation

It is well known that equivariance (2.31) implies that the nonlinear operator in (2.32) admits linear invariant subspaces, which we now outline. Let $\mathcal{X}$ denote configuration space, defined by $(2.13)_1$, viz.,

$$\mathcal{X}=\left(C^1([0,L),\mathbb{R}^3)\times C^1([0,L),S^3)\right)^3, \tag{2.33}$$

where $S^3$ is the set of all unit quaternions, and let $\mathcal{X}_o$ denote subset of all configurations satisfying the boundary conditions and joint-compatibility conditions, viz., $(2.4)_1$, $(2.5)_1$ and $(2.13)_{2,3.}$ Then $\mathbf{f}:\mathbb{R}\times\mathcal{X}_o\rightarrow\mathcal{Y}$, where the target space is the linear vector space $\mathcal{Y}=C([0,L],\mathbb{R}^{18})$. Let $\Sigma\subset C_{3v}$ be a subgroup of the tripod symmetry group $C_{3v}$, and define the $\Sigma$ fixed-point space

$$\mathcal{X}_o^{\Sigma}:=\{\mathbf{u}\in\mathcal{X}_o:\boldsymbol{T}_G\mathbf{u}=\mathbf{u}\text{ for all }\mathbf{G}\in\Sigma\}. \tag{2.34}$$

This is simply the set of all admissible configurations enjoying the symmetry characterized by the subgroup $\Sigma$. In any case, (2.31) implies

$$\mathbf{f}(\lambda,\cdot):\mathbb{R}\times\mathcal{X}_o^{\Sigma}\rightarrow\mathcal{Y}^{\Sigma}, \tag{2.35}$$

where $\mathcal{Y}^{\Sigma}$ is the target fixed-point space

$$\mathcal{Y}^{\Sigma}:=\{\mathbf{y}\in\mathcal{Y}:\tilde{\boldsymbol{T}}_G\mathbf{y}=\mathbf{y}\text{ for all }\mathbf{G}\in\Sigma\}.$$

Indeed, if $\mathbf{u}\in\mathcal{X}_o^{\Sigma}$, then (2.31) and (2.34) immediately give $\mathbf{f}(\lambda,\mathbf{u})=\mathbf{f}(\lambda,\boldsymbol{T}_G\mathbf{u})=\tilde{\boldsymbol{T}}_G\mathbf{f}(\lambda,\mathbf{u})$ for all $\mathbf{G}\in\Sigma$.

The main idea is to choose subgroups strategically, viz., in such a way that the symmetry-reduced bifurcation problem (2.32), (2.35) entails only a 1-dimensional bifurcation analysis, which is considerably simpler than the analysis of a multiple-bifurcation point. This is usually called the *equivariant branching lemma*. In particular, numerical branch switching is routine in the 1D case. We use generic, local-bifurcation arguments [9], [15] to guide our analyses before computing solution paths. The group $C_{3v}$ has three non-equivalent irreducible representations (irreps), denoted $\gamma_G^{(k)}, k=1,2,3,$ with generators

$$\begin{aligned}
&\gamma_{R_{2\pi/3}}^{(1)}=\gamma_E^{(1)}=1,\\
&\gamma_{R_{2\pi/3}}^{(2)}=-\gamma_E^{(2)}=1,\\
&\gamma_{R_{2\pi/3}}^{(3)}=\frac{1}{2}\begin{bmatrix}-1 & \sqrt{3}\\ -\sqrt{3} & -1\end{bmatrix},\ \gamma_E^{(3)}=\begin{bmatrix}1 & 0\\ 0 & -1\end{bmatrix},
\end{aligned} \tag{2.36}$$

respectively, cf. [17]. The first irrep (1) is always associated with symmetry-preserving configurations, while the other two provide a preview of the type of potential symmetry-breaking bifurcation modes we can expect.

Irrep (2) is associated with a potential 1-dimensional bifurcation, preserving rotational symmetry while breaking reflection symmetry. The latter conclusion follows from the observation that $\gamma_E^{(2)}=-1$. The single bifurcation equation – in the parameter $\lambda$ and the unknown amplitude – is odd in the latter variable. Consequently, the bifurcation will be a “pitchfork”.

Irrep (3) is associated with a potential double bifurcation point, i.e., the null space of the tangent stiffness is generically 2-dimensional. This is in consonance with the fact that representation (3) involves $2\times 2$ matrices. Unlike the previous case, there are now two local bifurcation equations – in the parameter $\lambda$ and two unknown amplitudes. Moreover, the bifurcation equations are equivariant under the matrix representation $\gamma^{(3)}$, the latter of which is the complete symmetry group of an equilateral triangle in the plane, denoted $D_3$. There are three equivalent ways to break the symmetry leaving that of an isosceles triangle. For instance, one of these has the symmetry group $\Sigma = \{\mathbf{I}, \mathbf{E}\}$. Accordingly, we expect three equivalent bifurcating branches. It is enough to compute only one; by virtue of (2.31), (2.32), the other two can be generated via consecutive actions of $(2.16)_1$. One such branch of solutions can be computed efficiently as a 1-dimensional bifurcation via the restriction (2.35) employing the simple reflection subgroup $\Sigma$ just mentioned. Finally, we mention that the equivariance of the bifurcation equations under $\gamma^{(3)}$ also implies that the bifurcation equations exhibit a quadratic term, and thus, a potential bifurcation will be transcritical in this case.

For the purpose of numerical computations, we choose the following quadratic, "diagonal" complementary energy function for each of the rods in the tripod:

$$\Gamma = \frac{L}{2}\left(\frac{n_1^2}{EA} + 2n_1 + \frac{n_2^2 + n_3^2}{GA_S} + \frac{m_1^2}{GJ} + \frac{m_2^2}{EI_2} + \frac{m_3^2}{EI_3}\right), \qquad (2.37)$$

with Young's modulus $E = 10^5\ Pa$, shear modulus $G = E/2$, cross-sectional area $A$, shear area $A_S$, torsional moment of inertia $J$, and bending moments of inertia $I_1, I_2$. We choose $L = 1$ for the rod lengths. Although this appears trivial here, we leave the multiplicative factor "$L$" in (2.37), anticipating structures for which the structural members need not have the same lengths, e.g., in the Section 3. We assume square cross-sections with geometric properties summarized below in Table 1.

| **Type of section** | **Width $w\ (m)$** | **Height $h\ (m)$** | **moment of inertia $I_2\ (m^4)$** | **moment of inertia $I_3\ (m^4)$** | **Torsional moment of inertia $J\ (m^4)$** | **Area** | **Shear area** |
|---|---|---|---|---|---|---|---|
| ***Square*** | $\frac{6}{34}$ | $\frac{6}{34}$ | $=\frac{w^3h}{12}$ $\approx 8.082\ 10^{-5}$ | $=\frac{wh^3}{12}$ | $0.141hw^3$ | $wh$ | $\frac{5wh}{6}$ |

*Table 1: Geometric characteristics of bar cross-sections*

We employ the path-following code AUTO [5], which detects and computes simple (i.e., 1-dimensional) bifurcations. As such, we begin with an analysis of the full tripod without symmetry restrictions in order capture the symmetry-preserving primary solution and any potential bifurcations associated with irrep $\gamma^{(2)}$. The field equations in local coordinates for each rod are given by (1.16) and (1.21), supplemented by (1.20) and (1.22). For each rod, we index all quantities appearing in the set of equations by a parenthetical superscript, viz., $\overline{r}^{(j)}, \overline{\mathrm{q}}^{(j)} = (q_o^{(j)}, \overline{q}^{(j)}), \underline{n}^{(j)}, \underline{m}^{(j)}, j = 1,2,3.$ Similarly, the compatibility equations $(2.4)_1$, $(2.13)_2$, the boundary conditions $(2.5)_1$, $(2.13)_3$, and the equilibrium conditions $(2.11)_2$ at joint A all need to be written with respect to the local coordinate system of each rod.

We first introduce

$$\mathbf{i}_k = \mathbf{Q}\mathbf{e}_k^{(1)},\ k = 1,2,3, \tag{2.38}$$

where $\mathbf{Q}$ is the rotation with matrix relative to the global basis given by

$$[Q] = \frac{1}{2}\begin{bmatrix} 1 & 0 & \sqrt{3} \\ 0 & 1 & 0 \\ -\sqrt{3} & 0 & 1 \end{bmatrix}, \tag{2.39}$$

The latter of which follows directly from $(2.1)_1$. By virtue of $(2.1)_2$ and (2.2), the compatibility equations $(2.4)_1$, $(2.13)_2$ become

$$\begin{aligned} &[Q]\overline{r}^{(1)}(0) = [R_{2\pi/3}][Q]\overline{r}^{(2)}(0) = [R_{4\pi/3}][Q]\overline{r}^{(3)}(0), \\ &q_o^{(1)}(0) = q_o^{(2)}(0) = q_o^{(3)}(0), \text{ and} \\ &[Q]\overline{q}^{(1)}(0) = [R_{2\pi/3}][Q]\overline{q}^{(2)}(0) = [R_{4\pi/3}][Q]\overline{q}^{(3)}(0), \end{aligned} \tag{2.40}$$

respectively. Similarly, the boundary conditions $(2.5)_1$, $(2.13)_3$ now read

$$\begin{aligned} &[Q]\overline{r}^{(1)}(1) = [Q]\overline{r}^{(2)}(1) = [Q]\overline{r}^{(3)}(1) = \begin{bmatrix} L/2 \\ 0 \\ 0 \end{bmatrix}, \\ &q_o^{(1)}(1) = q_o^{(2)}(1) = q_o^{(3)}(1) = 1, \text{ and} \\ &\overline{q}^{(1)}(1) = \overline{q}^{(2)}(1) = \overline{q}^{(3)}(1) = \overline{0}, \end{aligned} \tag{2.41}$$

respectively. For joint equilibrium, we first recall from Section 1 that $\mathbf{n}^{(j)} = n_k^{(j)}\mathbf{R}^{(j)}\mathbf{e}_k^{(j)}, j = 1,2,3,$ where $\mathbf{R}^{(j)} = R_{kl}^{(j)}\mathbf{e}_k^{(j)} \otimes \mathbf{e}_l^{(j)}$. Then by $(2.1)_2$ and (2.38), $(2.11)_2$ reads

$$\begin{aligned} &[Q]\overline{R}^{(1)}(\overline{\mathrm{q}}^{(1)}(0))\underline{n}^{(1)}(0) + [R_{2\pi/3}][Q]\overline{R}^{(2)}(\overline{\mathrm{q}}^{(2)}(0))\underline{n}^{(2)}(0) + [R_{4\pi/3}][Q]\overline{R}^{(3)}(\overline{\mathrm{q}}^{(3)}(0))\underline{n}^{(3)}(0) = \begin{bmatrix} 0 \\ 0 \\ \lambda \end{bmatrix}, \\ &[Q]\overline{R}^{(1)}(\overline{\mathrm{q}}^{(1)}(0))\underline{m}^{(1)}(0) + [R_{2\pi/3}][Q]\overline{R}^{(2)}(\overline{\mathrm{q}}^{(2)}(0))\underline{m}^{(2)}(0) + [R_{4\pi/3}][Q]\overline{R}^{(3)}(\overline{\mathrm{q}}^{(3)}(0))\underline{m}^{(3)}(0) = \underline{0}. \end{aligned} \tag{2.42}$$

The formulation from [13], outlined in Section 1, also requires the specification of (1.19) as a boundary condition whenever the latter does not involve a purely kinematic rotation. In view of $(2.40)_{2,3}$, it is enough to prescribe

$$(q_o^{(1)}(0))^2 + \overline{q}^{(1)}(0)\cdot\overline{q}^{(1)}(0) = 1. \tag{2.43}$$

We note that (2.41) represents 21 boundary conditions at $s = 1$. The joint compatibility equations (2.40) provide 14 more at $s = 0$, say, by relating $\overline{r}^{(\alpha)}(0),(q_o^{(\alpha)}(0),\overline{q}^{(\alpha)}(0))$ to $\overline{r}^{(1)}(0),(q_o^{(1)}(0),\overline{q}^{(1)}(0))$ for $\alpha = 2,3$. The joint equilibrium equations (2.42) together with (2.43) account for 7 more - for a total of 21 boundary conditions at $s = 0$. Altogether, we have 42 boundary conditions. This is precisely the number required for the boundary value problem comprising (1.16), (1.21) for each rod, totaling 39 first-order equations in 42 unknowns (including the three dummy parameters $\mu^{(j)}, j = 1,2,3$). Note that we could easily express the system as 42 equations by simply appending $\dot{\mu}^{(j)} = 0, j = 1,2,3$.

We now discuss the results obtained via AUTO for the full structure. The symmetry-preserving solutions obtained, associated with irrep $\gamma^{(1)}$, are depicted in Figure 2(a). Surprisingly, the results include not only the primary path but also a symmetry-preserving branch of bifurcating solutions.

Sample deformed configurations of the structure along each solution path are also given in Figure 2(b). Symmetry-preserving bifurcation is not "generic". Here the phenomenon arises on account of the employed constitutive law summarized in Table 1, which corresponds to a realistic model for engineering rods: The bending and twisting stiffnesses are several orders of magnitude smaller than the axial and shear stiffnesses, i.e., the rods are essentially inextensible and unshearable. As such, all three "buckle" simultaneously under compression, giving rise to the symmetry-preserving bifurcation. The same phenomenon is well documented for such a tripod comprising inextensible rods with pin-jointed connections [1].

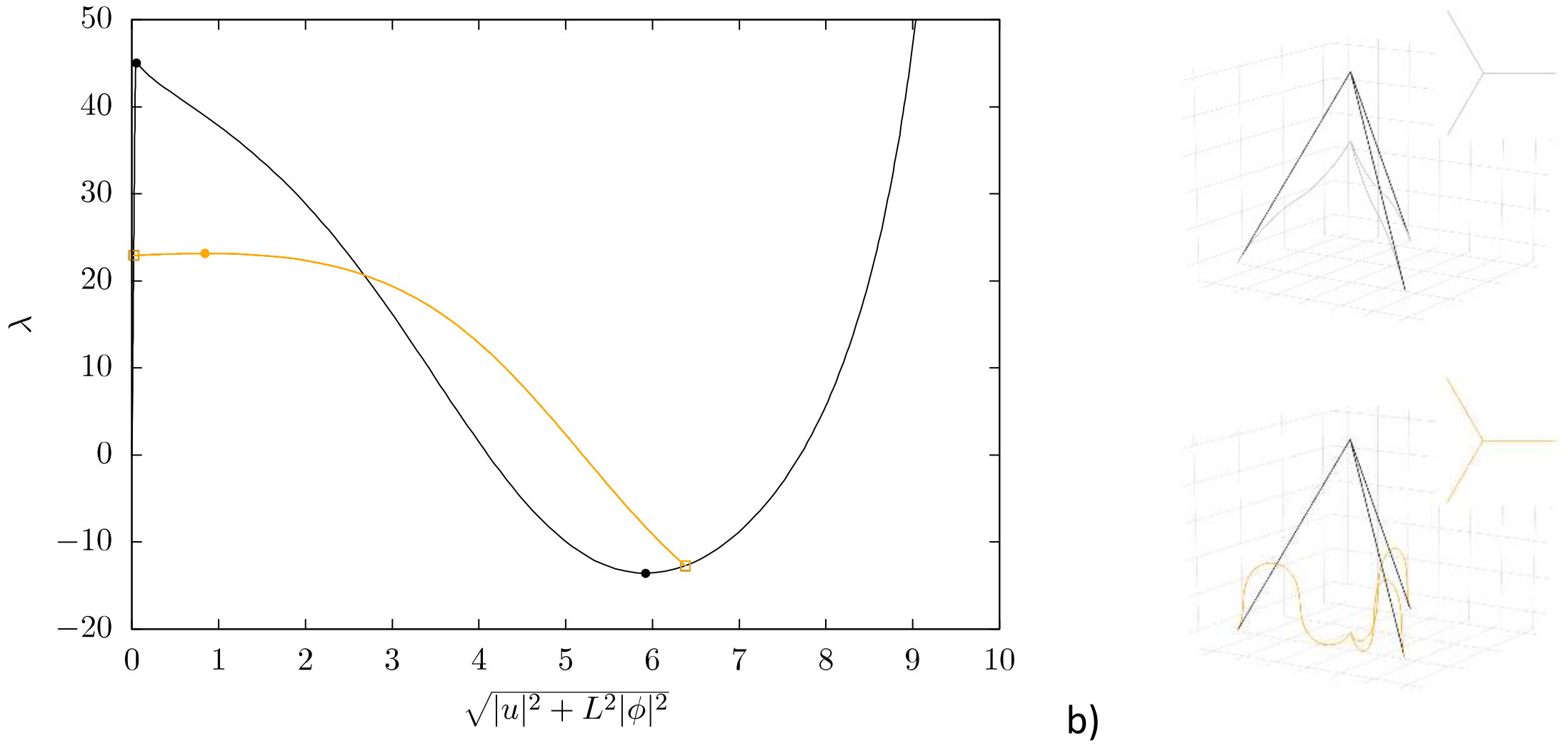

a) b)

*Figure 2. a) The principal path (black), bifurcation points and a bifurcated branch (orange) corresponding to symmetry-preserving bifurcation, each associated with the trivial 1D irrep (1). Empty square markers denote bifurcation points while full circles denote limit points. b) The corresponding deformed configurations (in grey and orange) plotted against the undeformed configuration (in black), perspective and top (insert) views.*

The analysis of the full structure via AUTO also reveals a symmetry-breaking bifurcation associated with irrep $\gamma^{(2)}$. The primary and bifurcating solution branches along with a sample configuration on the latter are depicted in Figure 3. The associated configurations maintain only $C_3$- symmetry. We observe that the bifurcation load occurs at a much smaller value than that of the symmetry-preserving bifurcation. We remark that the projected curve in Figure 3 appears to show a double intersection at the bifurcation point on the right. In fact, the flat, upper part of the curve shown there does not actually contain the bifurcation point. Another symmetry-breaking bifurcation of the same type occurring at a higher load level is also present on the principal path, which we do not pursue here.

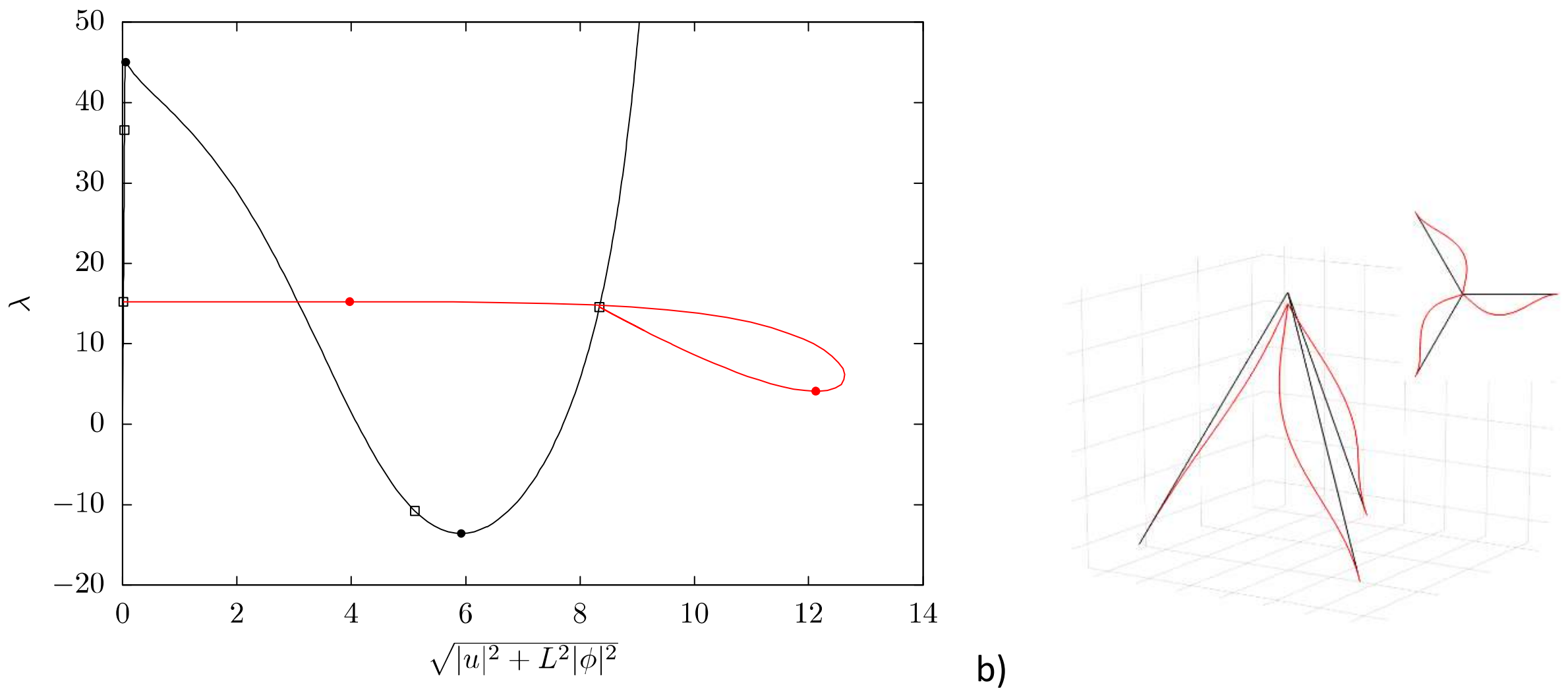

*Figure 3. a) The principal path (black), bifurcation points and first bifurcated branch (red) corresponding to bifurcations belonging to the non-trivial 1D irrep (2). Empty square markers denote bifurcation points while full circles denote limit points. b) The corresponding deformed configuration (in red) plotted against the undeformed configuration (in black), perspective and top (insert) views.*

Double bifurcation points, associated here with irrep $\gamma^{(3)}$, are not detected by the code AUTO. Accordingly, symmetry reduction via (2.35), rendering the bifurcation simple, is needed. As discussed above, we choose the subgroup $\Sigma = \{\mathbf{I}, \mathbf{E}\} \cong Z_2$, corresponding to reflection symmetry about the plane $span\{\mathbf{i}_1, \mathbf{i}_3\}$, cf. Figure 1. Our approach is facilitated by working with the substructure comprising rods 1 and 2, accompanied by reflection conditions imposed at the joint A. To begin, we note that the fixed boundary conditions (2.42) for rods 1 and 2 are unchanged.

In view of $(2.4)_1$ and $(2.13)_2$, we use $(2.13)_1$, $(2.16)_2$ and $(2.17)_2$, to deduce

$$\mathbf{E}\mathbf{r}_A = \mathbf{r}_A \text{ and } -\mathbf{E}\mathbf{q}_A = \mathbf{q}_A, \tag{2.44}$$

for the reduced problem (2.35). Then again by $(2.4)_1$ and $(2.13)_2$, we employ $(2.25)_1$ to find

$$r_2^{(1)}(0) = q_1^{(1)}(0) = q_3^{(1)}(0) = 0. \tag{2.45}$$

In keeping with the formulation of [13], we also require

$$(q_o^{(1)}(0))^2 + (q_2^{(1)}(0))^2 = 1. \tag{2.46}$$

The conditions for rod 2 follow similarly. From (2.1) and (2.38), it follows that $\mathbf{e}_\ell^{(2)} = \mathbf{R}_{2\pi/3}\mathbf{Q}\mathbf{i}_\ell, \ell = 1,2,3,$ and thus, $\mathbf{r}^{(2)} = r_k^{(2)}\mathbf{R}_{2\pi/3}\mathbf{Q}\mathbf{i}_k$. Then (2.25) and $(2.44)_1$ imply $\mathbf{i}_2 \cdot \mathbf{r}^{(2)}(0) = \mathbf{i}_2 \cdot [\mathbf{R}_{2\pi/3}\mathbf{Q}\mathbf{i}_k] r_k^{(2)}(0) = 0$, leading to

$$\sqrt{3} r_1^{(2)} + 2r_2^{(2)} + 3r_3^{(2)} = 0. \tag{2.47}$$

In the same manner using $(2.44)_2$, we obtain

$$2q_1^{(2)}(0) - \sqrt{3} q_2^{(2)}(0) = 0, \; 3q_2^{(2)}(0) - 2q_3^{(2)}(0) = 0. \tag{2.48}$$

Again, we enforce (1.19) at $s = 0$, viz.,

$$(q_o^{(2)}(0))^2 + 4(q_2^{(2)}(0))^2 = 1, \tag{2.49}$$

where we have made use of (2.48).

From $(2.29)_2$, the reduced problem (2.35) delivers $\mathbf{n}^{(3)} = \mathbf{E}\mathbf{n}^{(2)}$ and $\mathbf{m}^{(3)} = -\mathbf{E}\mathbf{m}^{(2)}$. The joint equilibrium equations $(2.11)_2$ then read

$$\mathbf{n}^{(1)}(0) + (\mathbf{I}+\mathbf{E})\mathbf{n}^{(2)}(0) = \lambda\mathbf{i}_3, \ \ \mathbf{m}^{(1)}(0) + (\mathbf{I}-\mathbf{E})\mathbf{m}^{(2)}(0) = \mathbf{0}. \tag{2.50}$$

Finally, we express (2.50) relative to the fixed global basis $\{\mathbf{i}_1, \mathbf{i}_2, \mathbf{i}_3\}$. Employing $(2.1)_2$, (2.14) and (2.39), we find

$$\begin{gathered}
[Q][\bar{R}^{(1)}(\bar{\mathsf{q}}^{(1)}(0))]\underline{n}^{(1)}(0) + ([I]+[E])[R_{2\pi/3}][Q][\bar{R}^{(2)}(\bar{\mathsf{q}}^{(2)}(0))]\underline{n}^{(2)}(0) = \begin{bmatrix} 0 \\ 0 \\ \lambda \end{bmatrix}, \\
[Q][\bar{R}^{(1)}(\bar{\mathsf{q}}^{(1)}(0))]\underline{m}^{(1)}(0) + ([I]-[E])[R_{2\pi/3}][Q][\bar{R}^{(2)}(\bar{\mathsf{q}}^{(2)}(0))]\underline{m}^{(2)}(0) = \underline{0}.
\end{gathered} \tag{2.51}$$

Note that (2.45)-(2.49) and (2.51) comprise 14 boundary conditions at $s = 0$. In addition, the fixed boundary conditions (2.42) for rods 1 and 2 specify 14 more at $s = 1$, for a total of 28 boundary conditions as required for the analysis of two rods. We employ AUTO for the reduced problem.

The numerical solution paths for $Z_2$-symmetric configurations are presented in Figure 4. As in the previous two cases, the principal, symmetry-preserving path is obtained, and the first bifurcating solution branch, breaking $C_{3v}$ to $Z_2$ symmetry, is depicted in Figure 4(a). As previously predicted, the bifurcation is transcritical (asymmetrical). Thus, the two "sides" of the bifurcating solution path are visible in the projected solution curve in Figure 4(a). Deformed configurations of each side of the bifurcating path are displayed in 4(b), where the blue shade of each is associated with that shade depicted in Figure 4(a). As discussed before there are two other bifurcating solution branches equivalent to the one reported here (say, generated by rotations). Another symmetry-breaking bifurcation of the same type but at a higher load level is also present on the principal path, which we do not pursue here.

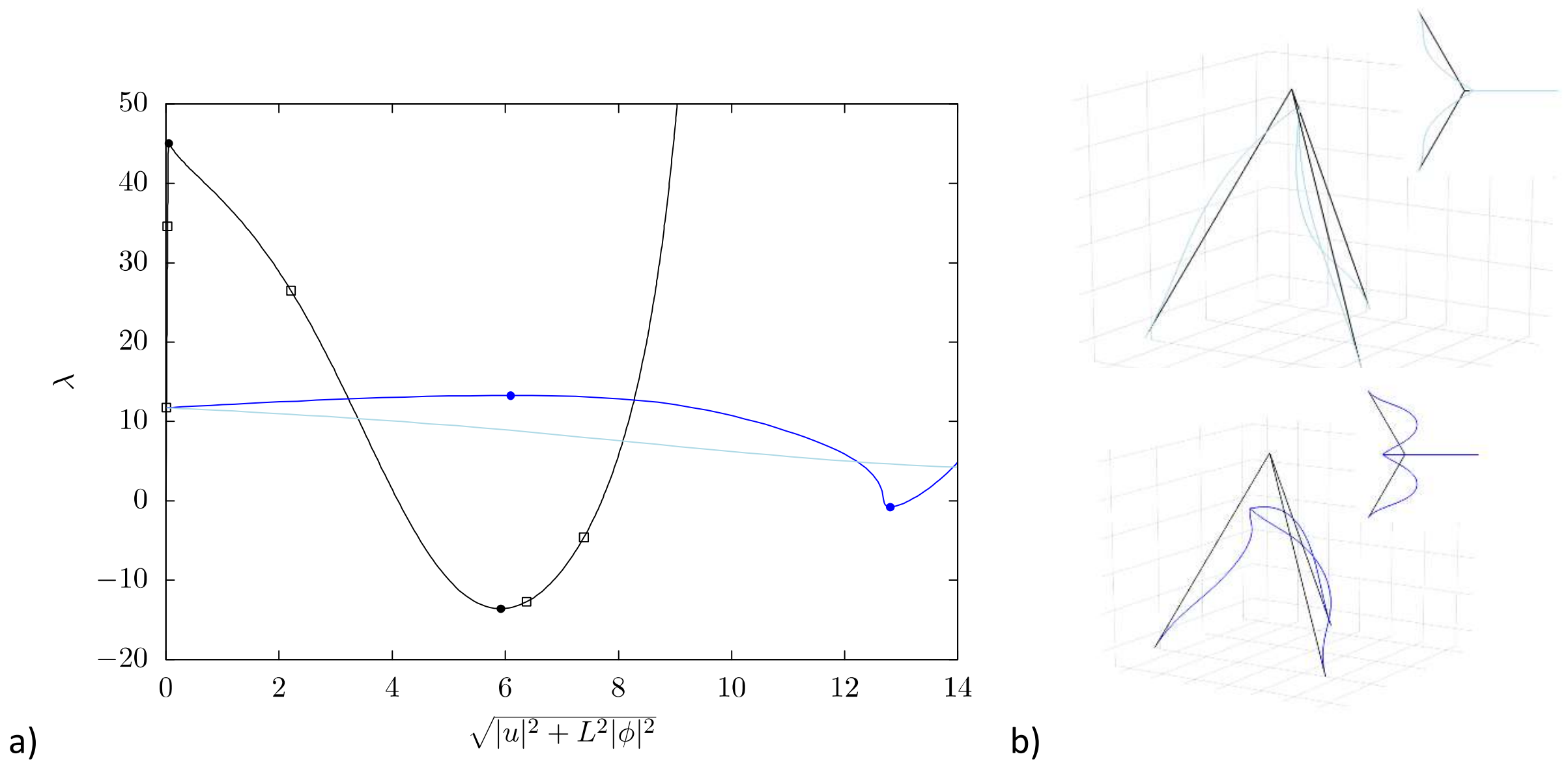

*Figure 4: (a) The principal path (black), bifurcation points and first bifurcated branches (blue) corresponding to bifurcation belonging to the 2D irrep (3). Empty square markers denote bifurcation points while full circles denote limit points. (b) The corresponding deformed configurations (in variations of blue) plotted against the undeformed configuration (in black), perspective and top (insert) views.*

Our results for the tripod are summarized in Figure 5. We note that the double bifurcation point, missed without symmetry reduction, is critical for this particular structure. The primary path is stable between $\lambda = 0$ and $\lambda = 12$, after which stability is lost. The transcritical bifurcating solution is also locally unstable, and the simple bifurcations previously discussed are all unstable.

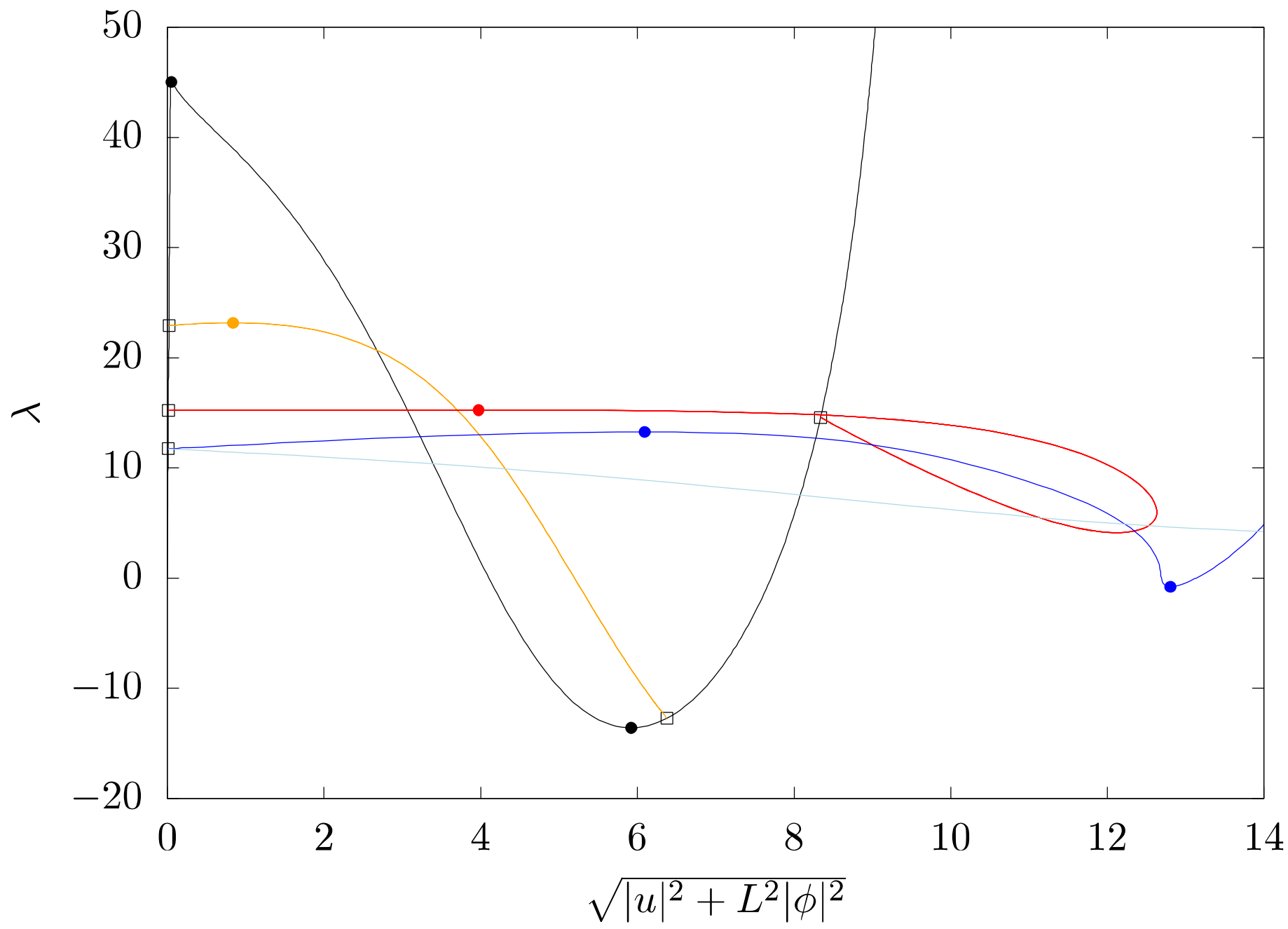

*Figure 5: Summary of primary bifurcating solution paths.*

## 3. A Space-Frame Dome

Next, we consider the hexagonal space-frame dome depicted below in Figure 6. We assume that each of the joints A-G is rigid or "welded". Each of the twenty-four rods comprising the dome is straight and prismatic with a reflection-symmetric cross-section with respect to planes perpendicular to span $\{\mathbf{i}_1, \mathbf{i}_2\}$. The material properties and cross-section of each rod are the same. Each of the joints A-G carries an external load of $-\lambda \mathbf{i}_3$. Each of the six supports of the structure is clamped. Accordingly, the system possesses the complete symmetry group

$$C_{6v} = \{\mathbf{R}_{k\pi/3}, \mathbf{ER}_{k\pi/3} : k = 1, 2, ..., 6\}, \tag{3.1}$$

where $\mathbf{E}$ is the reflection as defined in Section 1, and $\mathbf{R}_{\pi/3}$ is the proper, clockwise rotation about the global axis $\mathbf{i}_3$.

We employ the same methodology employed in Section 2 for the tripod. In this case, we require twenty-four placement and twenty-four unit-quaternion fields to describe a configuration. We designate the rods via the numbering scheme shown in Figure 6. Similar to (2.13), a configuration is uniquely specified by the fields

$$\mathbf{u} := \left(\mathbf{r}^{(1)}, \mathsf{q}^{(1)}, \mathbf{r}^{(2)}, \mathsf{q}^{(2)}, ..., \mathbf{r}^{(12)}, \mathsf{q}^{(12)};\ \mathbf{r}^{(13)}, \mathsf{q}^{(13)}, ..., \mathbf{r}^{(18)}, \mathsf{q}^{(18)};\ \mathbf{r}^{(19)}, \mathsf{q}^{(19)}, ..., \mathbf{r}^{(24)}, \mathsf{q}^{(24)}\right). \tag{3.2}$$

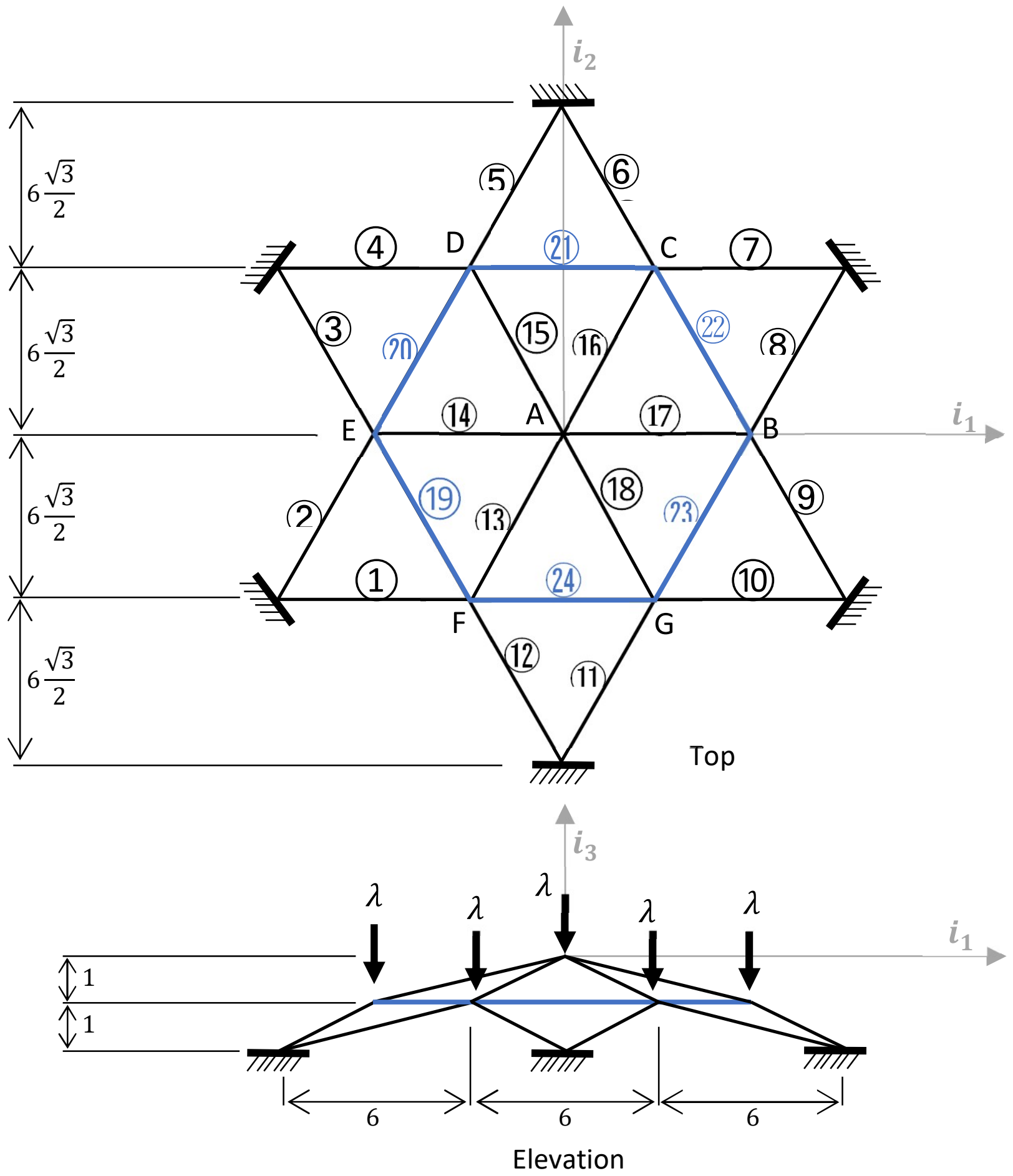

*Figure 6: Geometry, loading and Euclidian global basis for the hexagonal dome represented on top-down (up) and elevation(bottom) views. Colors represent bars of the same length length.*

Note that we group the deformation fields in (3.2) for rods of the same type together, with each group separated by semi-colons. The support rods appear in the first group, the top rods next, and the middle rods (shown in blue in Figure 6) last. The fields for the top and support rods are each defined over the interval $[0,\sqrt{37}]$, while each of the middle-rod fields are defined on $[0,6]$. We orient each rod, defining the local coordinate system, as follows: Each of the top rods "starts" at joint A. The middle rods are oriented clockwise about $\mathbf{i}_3$, and the support rods have their origins at the joints B-G, respectively.

The total potential energy of the structure is given by

$$\mathcal{V}[\lambda,\mathbf{u}] = \mathcal{U}[\mathbf{u}] + \lambda(\mathbf{r}_A + \mathbf{r}_B + ... + \mathbf{r}_G)\cdot\mathbf{i}_3, \tag{3.3}$$

where

$$\begin{aligned}&\mathcal{U}[\mathbf{u}] := \sum_{j=1}^{18}\int_0^{\sqrt{37}} W([\mathbf{R}^{(j)}]^T\dot{\mathbf{r}}^{(j)},[\mathbf{R}^{(j)}]^T\boldsymbol{\kappa}^{(j)})dx + \sum_{j=19}^{24}\int_0^{6} W([\mathbf{R}^{(j)}]^T\dot{\mathbf{r}}^{(j)},[\mathbf{R}^{(j)}]^T\boldsymbol{\kappa}^{(j)})dx,\\ &[\mathbf{R}^{(j)}]^T\dot{\mathbf{r}}^{(j)} = \nu_i^{(j)}\mathbf{e}_i^{(j)},\ [\mathbf{R}^{(j)}]^T\boldsymbol{\kappa}^{(j)} = \kappa_i^{(j)}\mathbf{e}_i^{(j)},\end{aligned}$$

and $W:(0,\infty)\times\mathbb{R}^5\to\mathbb{R}$ is the stored-energy function, common to each of the rods. We assume fixed boundary support conditions for members (1)-(12), similar to $(2.5)_1$ and $(2.13)_3$, and joint compatibility, as in $(2.4)_1$ and $(2.13)_{2,3}$. We give a sampling of both:

Boundary conditions:

$$\begin{aligned}&\mathbf{r}^{(11)}(\sqrt{37}) = \mathbf{r}^{(12)}(\sqrt{37}) = -6\sqrt{3}\mathbf{i}_2 - 2\mathbf{i}_3,\\ &\mathsf{q}^{(11)}(\sqrt{37}) = \mathsf{q}^{(12)}(\sqrt{37}) = (1,\mathbf{0}).\end{aligned} \tag{3.4}$$

Compatibility at joint A:

$$\begin{aligned}&\mathbf{r}^{(13)}(0) = \mathbf{r}^{(14)}(0) = ... = \mathbf{r}^{(18)}(0),\\ &\mathsf{q}^{(13)}(0) = \mathsf{q}^{(14)}(0) = ... = \mathsf{q}^{(18)}(0).\end{aligned} \tag{3.5}$$

The first-variation conditions for the potential energy deliver the equilibrium equations as in $(2.11)_1$:

$$\dot{\mathbf{n}}^{(j)} = \mathbf{0} \text{ and } \dot{\mathbf{m}}^{(j)} + \dot{\mathbf{r}}^{(j)}\times\mathbf{n}^{(j)} = \mathbf{0} \text{ in } (0,\sqrt{37}),\ \text{j=1,2,...,18; in } (0,6),\ \text{j=19,...,24}, \tag{3.6}$$

and joint equilibrium, as in $(2.11)_2$, at each of the joints A-G. For example, equilibrium at joint B reads

$$\begin{aligned}&\mathbf{n}^{(17)}(\sqrt{37}) + \mathbf{n}^{(22)}(6) + \mathbf{n}^{(23)}(0) + \mathbf{n}^{(8)}(0) + \mathbf{n}^{(9)}(0) = -\lambda\mathbf{i}_3,\\ &\mathbf{m}^{(17)}(\sqrt{37}) + \mathbf{m}^{(22)}(6) + \mathbf{m}^{(23)}(0) + \mathbf{m}^{(8)}(0) + \mathbf{m}^{(9)}(0) = \mathbf{0}.\end{aligned} \tag{3.7}$$

To properly describe the symmetry-group action on configuration space in terms of the two generators of $C_{6v}$, viz., $\mathbf{R}_{\pi/3}$ and $\mathbf{E}$, we need to account for the orientation of the continuous "ring" of rods 19-24. Let $\tau\in\mathbb{R}\bmod 36$ denote the continuous arc length of the ring, oriented clockwise starting at location B on rod 23, and let $\vartheta\in\mathbb{R}\bmod 2\pi$ denote the clockwise angle such that $\tau = 6\sin\vartheta/\sin(2\pi/3-\vartheta)$. We parametrize the ring via $\vartheta$ without changing the symbol of the dependent variable. Borrowing the representation for the elastic ring presented in [12], we then define

$$\boldsymbol{T}_{R_{\pi/3}}\mathbf{u} := (\mathbf{R}_{\pi/3}\mathbf{r}^{(12)}, \mathsf{R}_{\pi/3}\mathsf{q}^{(12)}, \mathbf{R}_{\pi/3}\mathbf{r}^{(1)}, \mathsf{R}_{\pi/3}\mathsf{q}^{(1)}, \mathbf{R}_{\pi/3}\mathbf{r}^{(2)}, \mathsf{R}_{\pi/3}\mathsf{q}^{(2)}, \ldots, \mathbf{R}_{\pi/3}\mathbf{r}^{(11)}, \mathsf{R}_{\pi/3}\mathsf{q}^{(11)};$$
$$\mathbf{R}_{\pi/3}\mathbf{r}^{(18)}, \mathsf{R}_{\pi/3}\mathsf{q}^{(18)}, \mathbf{R}_{\pi/3}\mathbf{r}^{(13)}, \mathsf{R}_{\pi/3}\mathsf{q}^{(13)}, \ldots, \mathbf{R}_{\pi/3}\mathbf{r}^{(17)}, \mathsf{R}_{\pi/3}\mathsf{q}^{(17)};$$
$$\mathbf{R}_{\pi/3}\mathbf{r}^{(24)}(\vartheta-\pi/3), \mathsf{R}_{\pi/3}\mathsf{q}^{(24)}(\vartheta-\pi/3), \mathbf{R}_{\pi/3}\mathbf{r}^{(19)}(\vartheta-\pi/3), \mathsf{R}_{\pi/3}\mathsf{q}^{(19)}(\vartheta-\pi/3), \ldots,$$
$$\mathbf{R}_{\pi/3}\mathbf{r}^{(23)}(\vartheta-\pi/3), \mathsf{R}_{\pi/3}\mathsf{q}^{(23)}(\vartheta-\pi/3)), \tag{3.8}$$

$$\boldsymbol{T}_{E}\mathbf{u} :== (\mathbf{E}\mathbf{r}^{(4)}, \mathsf{E}\mathsf{q}^{(4)}, \mathbf{E}\mathbf{r}^{(3)}, \mathsf{E}\mathsf{q}^{(3)}, \mathbf{E}\mathbf{r}^{(2)}, \mathsf{E}\mathsf{q}^{(2)}, \mathbf{E}\mathbf{r}^{(1)}, \mathsf{E}\mathsf{q}^{(1)}, \mathbf{E}\mathbf{r}^{(12)}, \mathsf{E}\mathsf{q}^{(12)},$$
$$\mathbf{E}\mathbf{r}^{(11)}, \mathsf{E}\mathsf{q}^{(11)}, \mathbf{E}\mathbf{r}^{(10)}, \mathsf{E}\mathsf{q}^{(10)}, \ldots, \mathbf{E}\mathbf{r}^{(6)}, \mathsf{E}\mathsf{q}^{(6)}, \mathbf{E}\mathbf{r}^{(5)}, \mathsf{E}\mathsf{q}^{(5)};$$
$$\mathbf{E}\mathbf{r}^{(15)}, \mathsf{E}\mathsf{q}^{(15)}, \mathbf{E}\mathbf{r}^{(14)}, \mathsf{E}\mathsf{q}^{(14)}, \mathbf{E}\mathbf{r}^{(13)}, \mathsf{E}\mathsf{q}^{(13)}, \mathbf{E}\mathbf{r}^{(18)}, \mathsf{E}\mathsf{q}^{(18)}, \mathbf{E}\mathbf{r}^{(17)}, \mathsf{E}\mathsf{q}^{(17)}, \mathbf{E}\mathbf{r}^{(16)}, \mathsf{E}\mathsf{q}^{(16)};$$
$$\mathbf{E}\mathbf{r}^{(20)}(-\vartheta), \mathsf{E}\mathsf{q}^{(20)}(-\vartheta), \mathbf{E}\mathbf{r}^{(19)}(-\vartheta), \mathsf{E}\mathsf{q}^{(19)}(-\vartheta), \mathbf{E}\mathbf{r}^{(24)}(-\vartheta), \mathsf{E}\mathsf{q}^{(24)}(-\vartheta),$$
$$\mathbf{E}\mathbf{r}^{(23)}(-\vartheta), \mathsf{E}\mathsf{q}^{(23)}(-\vartheta), \mathbf{E}\mathbf{r}^{(22)}(-\vartheta), \mathsf{E}\mathsf{q}^{(22)}(-\vartheta), \mathbf{E}\mathbf{r}^{(21)}(-\vartheta), \mathsf{E}\mathsf{q}^{(21)}(-\vartheta)).$$

Observe that the reflection reverses the orientations of rods 19-24 in this case, which explains the appearance of "$-\vartheta$" in the arguments of those actions. The group actions on the support rods and the top rods do not affect orientations and thus do not act on the independent variable. As in Section 2, the total potential energy (3.3) is invariant under the group action, viz.,

$$\mathcal{V}[\lambda, \boldsymbol{T}_G\mathbf{u}] = \mathcal{V}[\lambda, \mathbf{u}] \text{ for all } \mathbf{G} \in C_{6v}. \tag{3.9}$$

Although we omit the tedious details, we observe that the invariance of the internal energy $\mathcal{U}$ is a consequence of the following: Given a configuration field $\mathbf{u}$ as in (3.2), then $\boldsymbol{T}_G\mathbf{u}$, $\mathbf{G} \in C_{6v}$, can be interpreted as either a rotation by an integer multiple of $\pi/3$ or a reflection of the configuration across a plane perpendicular to $span\{\mathbf{i}_1, \mathbf{i}_2\}$ making an angle of an integer multiple of $\pi/6$ with respect to $span\{\mathbf{i}_1, \mathbf{i}_3\}$. In each of the three levels of the structure, the rods are all identical and have reflection symmetry in their cross-sections. Thus, the internal energy is invariant. For the loading term in (3.3), we deduce from the generators (3.8) that

$$\mathbf{r}_A + \mathbf{r}_B + \ldots + \mathbf{r}_G \to \mathbf{R}_{\pi/3}(\mathbf{r}_A + \mathbf{r}_B + \ldots + \mathbf{r}_G)$$

and

$$\mathbf{r}_A + \mathbf{r}_B + \ldots + \mathbf{r}_G \to \mathbf{E}(\mathbf{r}_A + \mathbf{r}_B + \ldots + \mathbf{r}_G),$$

respectively. Of course, $\mathbf{R}_{\pi/3}^T\mathbf{i}_3 = \mathbf{E}\mathbf{i}_3 = \mathbf{i}_3$, and the invariance follows.

As in Section 2, the first variation of the potential energy and integration by parts deliver the strong form of the equilibrium equations, again denoted abstractly by (2.32). Moreover, the equivariance of the equilibrium equations follows from (3.9) as in Proposition 2.2, viz.,

$$\mathbf{f}(\lambda, \boldsymbol{T}_G\mathbf{u}) = \tilde{\boldsymbol{T}}_G\mathbf{f}(\lambda, \mathbf{u}) \text{ for all } \mathbf{G} \in C_{6v}, \tag{3.10}$$

where the generators of the action $\tilde{\boldsymbol{T}}_G$ in (3.10) are defined as follows: For

$$\boldsymbol{\sigma} := \left(\mathbf{n}^{(1)}, \mathbf{m}^{(1)}, \mathbf{n}^{(2)}, \mathbf{m}^{(2)}, \ldots, \mathbf{n}^{(12)}, \mathbf{m}^{(12)};\ \mathbf{n}^{(13)}, \mathbf{m}^{(13)}, \ldots, \mathbf{n}^{(18)}, \mathbf{m}^{(18)};\ \mathbf{n}^{(19)}, \mathbf{m}^{(19)}, \ldots, \mathbf{n}^{(24)}, \mathbf{m}^{(24)}\right),$$

$$\tilde{\boldsymbol{T}}_{R_{\pi/3}}\boldsymbol{\sigma} := (\mathbf{R}_{\pi/3}\mathbf{n}^{(12)}, \mathbf{R}_{\pi/3}\mathbf{m}^{(12)}, \mathbf{R}_{\pi/3}\mathbf{n}^{(1)}, \mathbf{R}_{\pi/3}\mathbf{m}^{(1)}, \mathbf{R}_{\pi/3}\mathbf{n}^{(2)}, \mathbf{R}_{\pi/3}\mathbf{m}^{(2)}, ..., \mathbf{R}_{\pi/3}\mathbf{n}^{(11)}, \mathbf{R}_{\pi/3}\mathbf{m}^{(11)};$$
$$\mathbf{R}_{\pi/3}\mathbf{n}^{(18)}, \mathbf{R}_{\pi/3}\mathbf{m}^{(18)}, \mathbf{R}_{\pi/3}\mathbf{n}^{(13)}, \mathbf{R}_{\pi/3}\mathbf{m}^{(13)}, ..., \mathbf{R}_{\pi/3}\mathbf{n}^{(17)}, \mathbf{R}_{\pi/3}\mathbf{m}^{(17)};$$
$$\mathbf{R}_{\pi/3}\mathbf{n}^{(24)}(\vartheta - \pi/3), \mathbf{R}_{\pi/3}\mathbf{m}^{(24)}(\vartheta - \pi/3), \mathbf{R}_{\pi/3}\mathbf{n}^{(19)}(\vartheta - \pi/3), \mathbf{R}_{\pi/3}\mathbf{m}^{(19)}(\vartheta - \pi/3), ...,$$
$$\mathbf{R}_{\pi/3}\mathbf{n}^{(23)}(\vartheta - \pi/3), \mathbf{R}_{\pi/3}\mathbf{m}^{(23)}(\vartheta - \pi/3)), \tag{3.11}$$

$$\tilde{\boldsymbol{T}}_{E}\boldsymbol{\sigma} :== (\mathbf{E}\mathbf{n}^{(4)}, -\mathbf{E}\mathbf{m}^{(4)}, \mathbf{E}\mathbf{n}^{(3)}, -\mathbf{E}\mathbf{m}^{(3)}, \mathbf{E}\mathbf{n}^{(2)}, -\mathbf{E}\mathbf{m}^{(2)}, \mathbf{E}\mathbf{n}^{(1)}, -\mathbf{E}\mathbf{m}^{(1)}, \mathbf{E}\mathbf{n}^{(12)}, -\mathbf{E}\mathbf{m}^{(12)},$$
$$\mathbf{E}\mathbf{n}^{(11)}, -\mathbf{E}\mathbf{m}^{(11)}, \mathbf{E}\mathbf{n}^{(10)}, -\mathbf{E}\mathbf{m}^{(10)}, ..., \mathbf{E}\mathbf{n}^{(6)}, -\mathbf{E}\mathbf{m}^{(6)}, \mathbf{E}\mathbf{n}^{(5)}, -\mathbf{E}\mathbf{m}^{(5)};$$
$$\mathbf{E}\mathbf{n}^{(15)}, -\mathbf{E}\mathbf{m}^{(15)}, \mathbf{E}\mathbf{n}^{(14)}, -\mathbf{E}\mathbf{m}^{(14)}, \mathbf{E}\mathbf{n}^{(13)}, -\mathbf{E}\mathbf{m}^{(13)}, \mathbf{E}\mathbf{n}^{(18)}, -\mathbf{E}\mathbf{m}^{(18)}, \mathbf{E}\mathbf{n}^{(17)}, -\mathbf{E}\mathbf{m}^{(17)}, \mathbf{E}\mathbf{n}^{(16)}, -\mathbf{E}\mathbf{m}^{(16)};$$
$$\mathbf{E}\mathbf{n}^{(20)}(-\vartheta), -\mathbf{E}\mathbf{m}^{(20)}(-\vartheta), \mathbf{E}\mathbf{n}^{(19)}(-\vartheta), -\mathbf{E}\mathbf{m}^{(19)}(-\vartheta), \mathbf{E}\mathbf{n}^{(24)}(-\vartheta), -\mathbf{E}\mathbf{m}^{(24)}(-\vartheta),$$
$$\mathbf{E}\mathbf{n}^{(23)}(-\vartheta), -\mathbf{E}\mathbf{m}^{(23)}(-\vartheta), \mathbf{E}\mathbf{n}^{(22)}(-\vartheta), -\mathbf{E}\mathbf{m}^{(22)}(-\vartheta), \mathbf{E}\mathbf{n}^{(21)}(-\vartheta), -\mathbf{E}\mathbf{m}^{(21)}(-\vartheta)).$$

The symmetry group $C_{6v}$ has six irreducible representations [17], the generators of which are:

$$\begin{aligned}
&\gamma^{(1)}_{R_{\pi/3}} = \gamma^{(1)}_{E} = 1, \\
&\gamma^{(2)}_{R_{\pi/3}} = -\gamma^{(2)}_{E} = 1, \\
&\gamma^{(3)}_{R_{\pi/3}} = -\gamma^{(3)}_{E} = -1, \\
&\gamma^{(4)}_{R_{\pi/3}} = \gamma^{(4)}_{E} = -1, \\
&\gamma^{(5)}_{R_{\pi/3}} = \frac{1}{2}\begin{bmatrix} -1 & \sqrt{3} \\ -\sqrt{3} & -1 \end{bmatrix}, \ \gamma^{(5)}_{E} = \begin{bmatrix} 1 & 0 \\ 0 & -1 \end{bmatrix}, \\
&\gamma^{(6)}_{R_{\pi/3}} = \frac{1}{2}\begin{bmatrix} 1 & \sqrt{3} \\ -\sqrt{3} & 1 \end{bmatrix}, \ \gamma^{(6)}_{E} = \begin{bmatrix} 1 & 0 \\ 0 & -1 \end{bmatrix}.
\end{aligned} \tag{3.12}$$

As in Section2, the first irrep $\gamma^{(1)}$ is associated with symmetry-preserving configurations. The 1-dimensional irreps (2)-(4) are each associated with potential 1-dimensional symmetry-breaking bifurcations. As before, the second irrep $\gamma^{(2)}$ is associated with configurations having pure rotational symmetry without reflections symmetry, viz., $C_6$- symmetry. The other two irreps $\gamma^{(3)}$ and $\gamma^{(4)}$ are associated with configurations having only $C_{3v}$- symmetry (with different orientations). This can be deduced from the observations $\gamma^{(3)}_{R_{2\pi/3}} = \gamma^{(3)}_{E} = 1$ and $\gamma^{(4)}_{R_{2\pi/3}} = \gamma^{(4)}_{E}\gamma^{(4)}_{R_{\pi/3}} = \gamma^{(4)}_{ER_{\pi/3}} = 1$. In each case, these are the generators of a 1-dimensional representation of $D_3 \cong C_{3v}$. In the potential case of bifurcation associated with the irreps $(3.12)_{2\text{-}4}$, we expect pitchforks in each case. This follows from the equivariance of the bifurcation equation under the representation, which induces oddness in the amplitude variable. As noted before in Section 2, the software AUTO detects and computes such simple bifurcations – in particular, without symmetry reduction. Again, we take a practical approach here and analyze to the full structure. For each rod in the framework, we assume the material properties summarized in Table 2.1, and we rescale according to (2.37), where $L$ denotes the length of the particular rod. In this way, we obtain a system of equations (1.16), (1.21) for each rod, all defined over the unit interval, the form of which is necessary for the use of AUTO. The rod equilibrium equations are combined with joint compatibility equations and joint equilibrium equations, samples of which are provided in (3.5) and (3.7), respectively. As in the case of the tripod of Section 2, the field equations for each rod need to be expressed with respect to its local basis, the details of which we omit here. In this case, we also employ the modified bifurcation criterion proposed in [8], enabling the code to work accurately for very large boundary-value problems.

The results for the full structure (no symmetry reductions) are now presented. The symmetry-preserving primary solution branch is depicted in Figure 7, and a few sample configurations corresponding to various solutions along it are provided in Figure 8. In this case, there are no symmetry-preserving bifurcations. In particular, the solution curve does not have self-intersections, which only arise in the projection shown in Figure 7. Several symmetry breaking bifurcations are indicated in Figure 7. We discuss only those occurring on the first "ascending" part of the solution path. The first at $\lambda = 1.15$ is a subcritical pitchfork. The global solution path is shown in Figure 9 along with a typical configuration. In particular, we observe that solutions on the bifurcating branch maintain $C_{3v}$-symmetry associated with the irrep $\gamma^{(4)}$. The latter follows from the observation that the configurations possess a reflection symmetry corresponding to $\boldsymbol{T}_E \boldsymbol{T}_{R_{\pi/3}} = \boldsymbol{T}_{ER_{\pi/3}}$. Remarkably the second solution branch, bifurcating at $\lambda = 1.22$, is of the same type, viz., a $C_{3v}$-symmetric subcritical pitchfork also associated with $\gamma^{(4)}$. The global solution path and a typical configuration are shown in Figure 10. Finally, the third solution path, bifurcating at $\lambda = 1.31$, and a typical configuration are given in Figure 11. These configurations also maintain $C_{3v}$- symmetry associated with irrep $\gamma^{(3)}$; the configurations possess a reflection symmetry corresponding to $\boldsymbol{T}_E$.

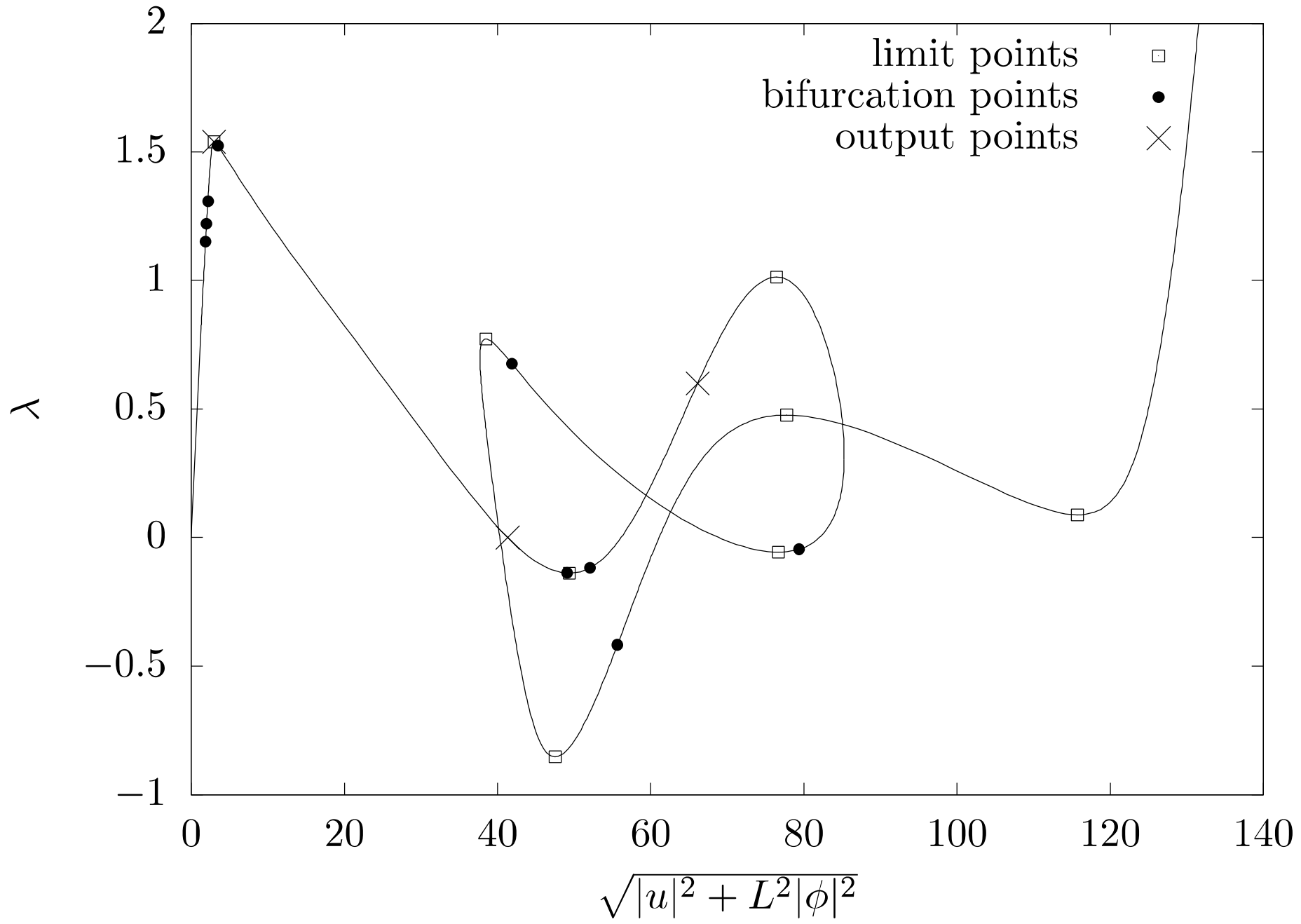

*Figure 7. Primary solution branch for the hexagonal dome.*

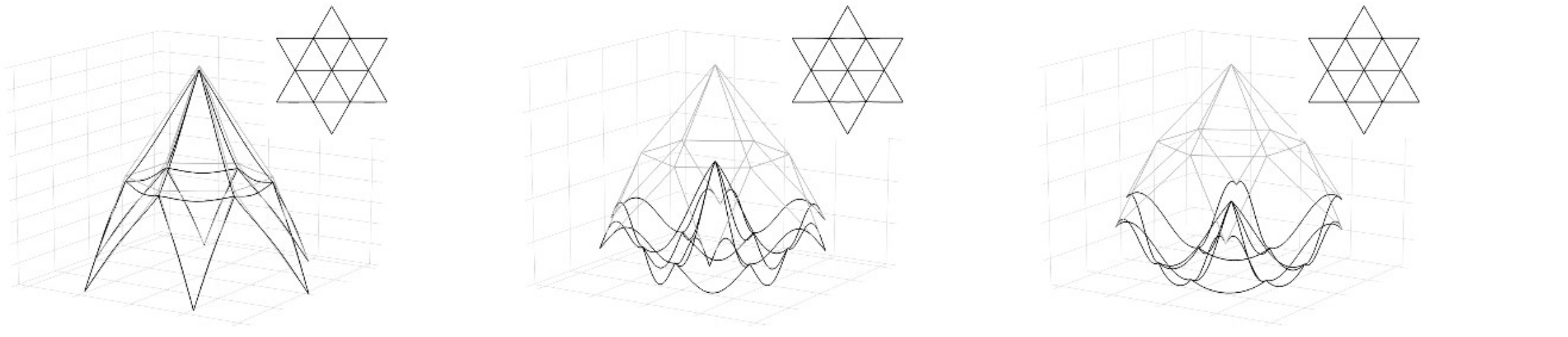

*Figure 8. Symmetry-preserving configurations along the primary path.*

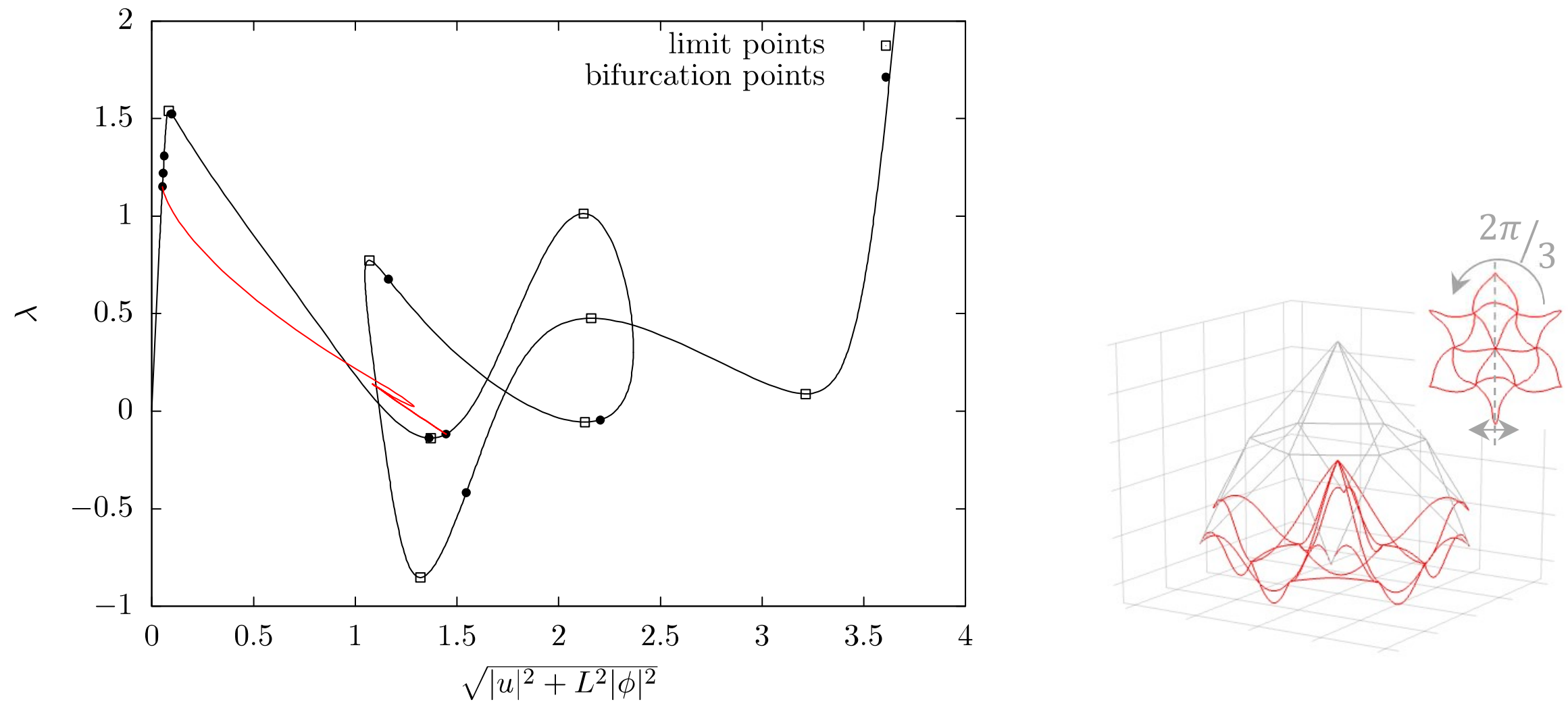

*Figure 9. Primary solution path and first $C_{3v}$-symmetric pitchfork bifurcation at* $\lambda =$ 1.15 *and a typical configuration.*

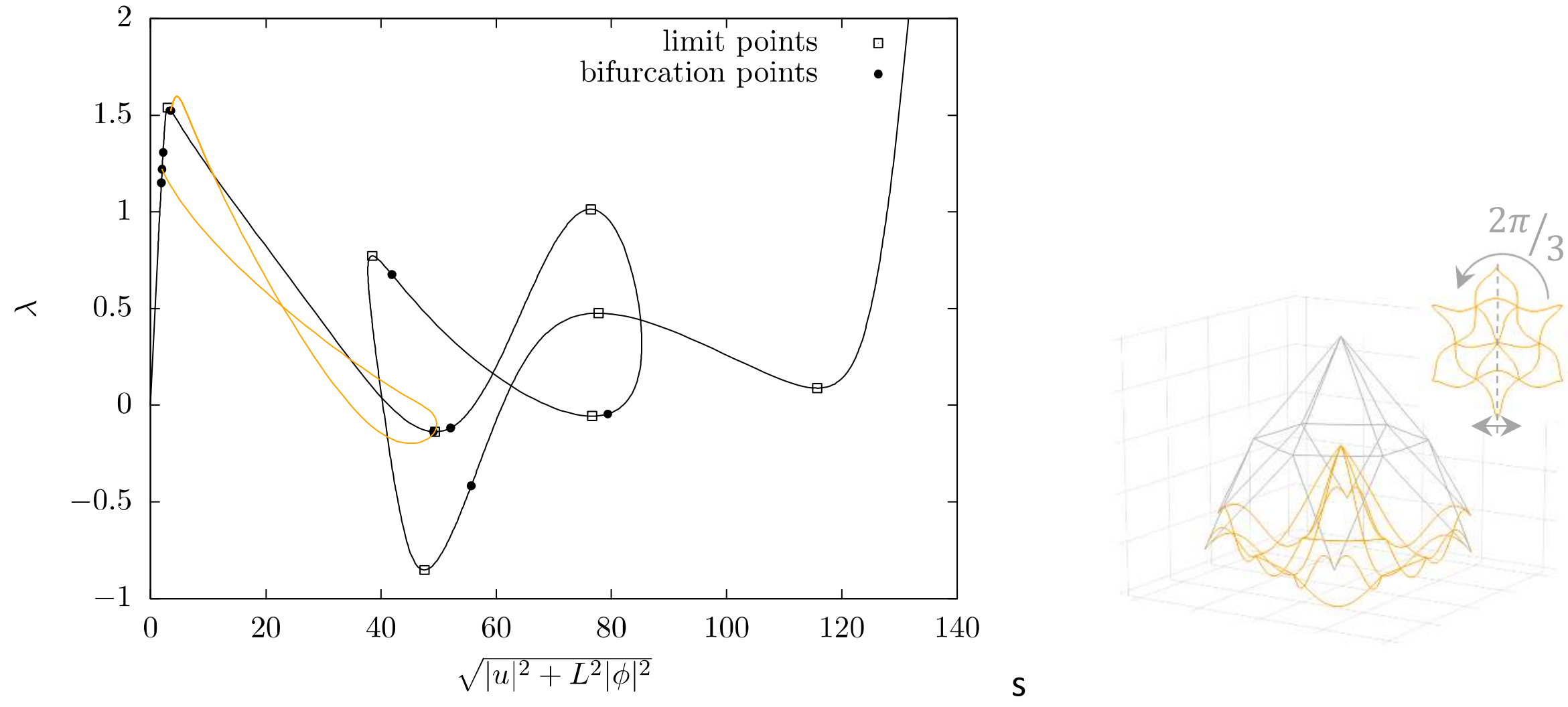

*Figure 10. Primary solution path and second $C_{3v}$-symmetric pitchfork bifurcation at* $\lambda =$ 1.22 *and a typical configuration.*

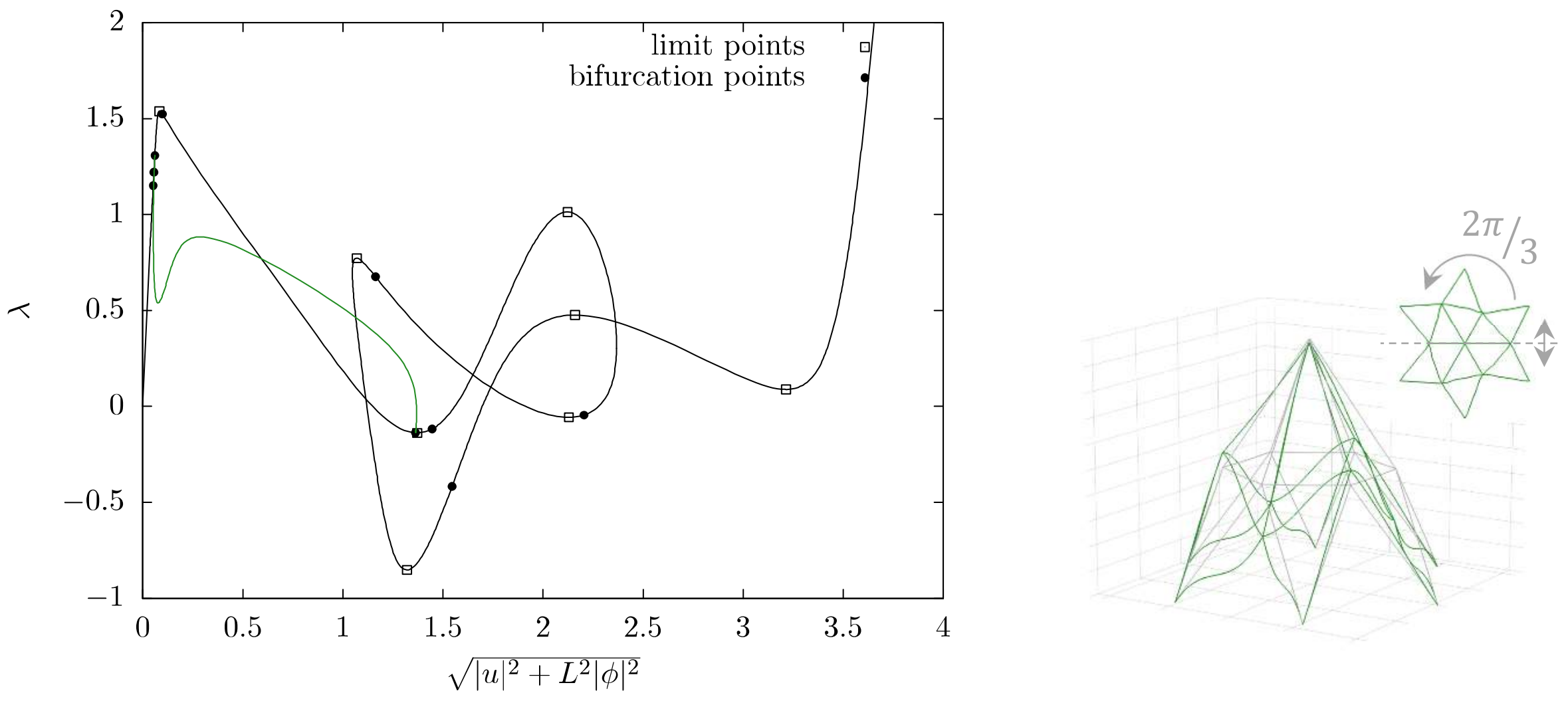

*Figure 11. Primary solution path and third $C_{3v}$-symmetric pitchfork bifurcation at* $\lambda = 1.31$ *and a typical configuration.*

Double bifurcation points for the complete structure, potentially associated here with the 2-dimensional irreps $\gamma^{(5)}$ and $\gamma^{(6)}$, cannot be detected directly by AUTO; symmetry reduction is required. The irrep $\gamma^{(6)}$ is a faithful representation of $D_6 (\cong C_{6v})$ on $\mathbb{R}^2$; $\gamma^{(5)}$ is not. But a comparison of the latter with (2.36)$_3$ reveals that $\gamma^{(5)}$ is associated (faithfully) with the symmetry $D_3 (\cong C_{3v})$ on $\mathbb{R}^2$, but with each element of the group appearing twice. As discussed in Section 2, there are three equivalent ways to break $D_3$-symmetry (to that of an isosceles triangle) leaving reflection symmetry. Accordingly, if there is bifurcation associated with $\gamma^{(5)}$, we expect three equivalent bifurcating branches; it is enough to compute only one of these. As discussed in Section 2, we expect such a bifurcation to be transcritical.

Symmetry-breaking bifurcation associated with $\gamma^{(6)}$ is more interesting: There are two *nonequivalent* ways to break $D_6$ symmetry (that of a hexagon) leaving reflection symmetry. One of the symmetry subgroups comprises the $2\times 2$ identity matrix and $\gamma_E^{(6)}$, while the other consists of the identity and $\gamma_E^{(6)}\gamma_{R_\pi}^{(6)} = \gamma_{ER_\pi}^{(6)} = \begin{bmatrix} -1 & 0 \\ 0 & 1 \end{bmatrix}$. Notice that the former has a reflection symmetry about a line connecting two opposing vertices of the hexagon, while the latter has a reflection symmetry about that connecting two opposing edges of the hexagon. Each of these, in turn, can be accomplished in two other equivalent ways. Thus, if there is bifurcation associated with $\gamma^{(6)}$, then we expect two non-equivalent bifurcating branches, each associated with two other equivalent branches – for a total of six bifurcating branches. Any such bifurcation is expected to be a pitchfork; the equivariance of the bifurcation equations under $\gamma^{(6)}$ implies oddness in the amplitude of the reduced 1D bifurcation equation [8].

To capture potential double bifurcations here (via AUTO), we require reduced problems (2.35) such that the bifurcations become simple, as in the $Z_2$- preserving solutions of Section 2. The main task is to obtain substructures (and boundary conditions) that admit only simple bifurcations. As in Section 2, we restrict the symmetry reductions only at the joints A-G. Based on the observations discussed in the paragraph above, it is apparent that substructures maintaining only reflection symmetry are required. The first substructure we employ is shown in Figure 12.

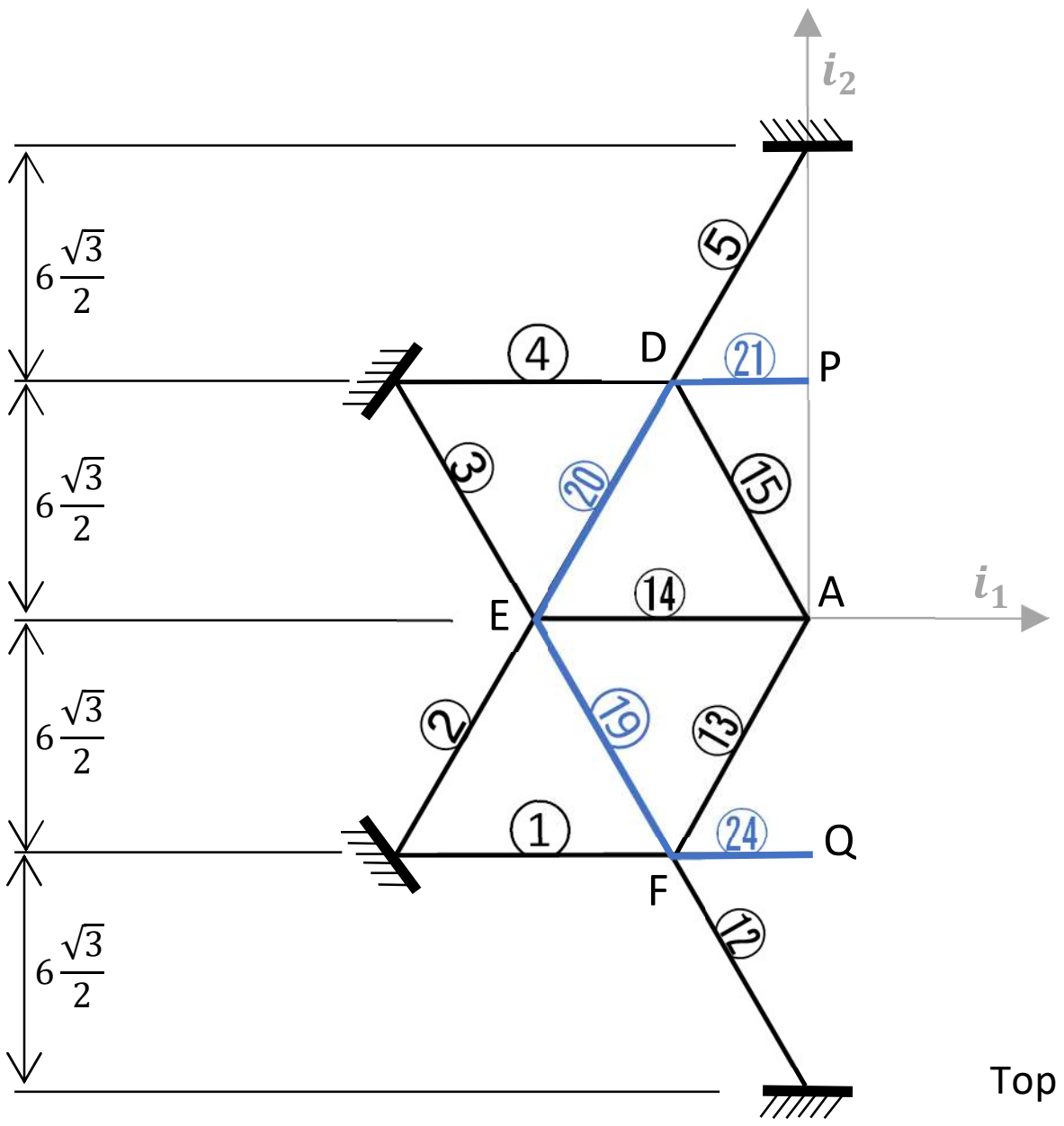

*Figure 12. Top view of the substructure associated with symmetry group* $\Sigma = \{\mathbf{I}, \mathbf{ER}_\pi\}$.

The boundary conditions at the supports and the joint compatibility and equilibrium conditions at joints D, E and F are the same as before (for the complete structure). The boundary conditions at joint A and at the newly exposed points P,Q arise from the reflection symmetry associated with the reduced problem. In view of (3.1), recall that $\mathbf{K} := \mathbf{ER}_\pi \in C_{6v}$ is a reflection across the plane $span\{\mathbf{i}_2, \mathbf{i}_3\}$. It is not hard to see that $\boldsymbol{T}_{\mathrm{K}} = \boldsymbol{T}_{ER_\pi} = \boldsymbol{T}_E \boldsymbol{T}_{R_\pi}$, and from (3.8), we find

$$\begin{aligned}\boldsymbol{T}_{\mathrm{K}}\mathbf{u} :== (&\mathbf{Kr}^{(10)}, \mathsf{Kq}^{(10)}, \mathbf{Kr}^{(9)}, \mathsf{Kq}^{(9)}, \mathbf{Kr}^{(8)}, \mathsf{Kq}^{(8)}, \mathbf{Kr}^{(7)}, \mathsf{Kq}^{(7)}, \mathbf{Kr}^{(6)}, \mathsf{Kq}^{(6)}, \mathbf{Kr}^{(5)}, \mathsf{Kq}^{(5)},\\ &\mathbf{Kr}^{(4)}, \mathsf{Kq}^{(4)}, \ldots, \mathbf{Kr}^{(1)}, \mathsf{Kq}^{(1)}, \mathbf{Kr}^{(12)}, \mathsf{Kq}^{(12)}, \mathbf{Kr}^{(11)}, \mathsf{Kq}^{(11)};\\ &\mathbf{Kr}^{(18)}, \mathsf{Kq}^{(18)}, \mathbf{Kr}^{(17)}, \mathsf{Kq}^{(17)}, \mathbf{Kr}^{(16)}, \mathsf{Kq}^{(16)}, \ldots, \mathbf{Kr}^{(13)}, \mathsf{Kq}^{(13)};\\ &\mathbf{Kr}^{(23)}(\pi-\vartheta), \mathsf{Kq}^{(23)}(\pi-\vartheta), \mathbf{Kr}^{(22)}(\pi-\vartheta), \mathsf{Kq}^{(22)}(\pi-\vartheta), \mathbf{Kr}^{(21)}(\pi-\vartheta), \mathsf{Kq}^{(21)}(\pi-\vartheta),\\ &\mathbf{Kr}^{(20)}(\pi-\vartheta), \mathsf{Kq}^{(20)}(\pi-\vartheta), \mathbf{Kr}^{(19)}(\pi-\vartheta), \mathsf{Kq}^{(19)}(\pi-\vartheta), \mathbf{Kr}^{(24)}(\pi-\vartheta), \mathsf{Kq}^{(24)}(\pi-\vartheta)),\end{aligned} \tag{3.13}$$

where $\mathsf{Kq} := (q_o, -\mathbf{Kq})$. Similarly, we have

$$\begin{aligned}\tilde{\boldsymbol{T}}_{\mathrm{K}}\boldsymbol{\sigma} := (&\mathbf{Kn}^{(10)}, -\mathbf{Km}^{(10)}, \mathbf{Kn}^{(9)}, -\mathbf{Km}^{(9)}, \mathbf{Kn}^{(8)}, -\mathbf{Km}^{(8)}, \mathbf{Kn}^{(7)}, -\mathbf{Km}^{(7)}, \ldots,\\ &\mathbf{Kn}^{(1)}, -\mathbf{Km}^{(1)}, \mathbf{Kn}^{(12)}, -\mathbf{Km}^{(12)}, \mathbf{Kn}^{(11)}, -\mathbf{Km}^{(11)};\\ &\mathbf{Kn}^{(18)}, -\mathbf{Km}^{(18)}, \mathbf{Kn}^{(17)}, -\mathbf{Km}^{(17)}, \ldots, \mathbf{Kn}^{(13)}, -\mathbf{Km}^{(13)};\\ &\mathbf{Kn}^{(23)}(\pi-\vartheta), -\mathbf{Km}^{(23)}(\pi-\vartheta), \mathbf{Kn}^{(22)}(\pi-\vartheta), -\mathbf{Km}^{(22)}(\pi-\vartheta),\\ &\mathbf{Kn}^{(21)}(\pi-\vartheta), -\mathbf{Km}^{(21)}(\pi-\vartheta), \mathbf{Kn}^{(20)}(\pi-\vartheta), -\mathbf{Km}^{(20)}(\pi-\vartheta),\\ &\mathbf{Kn}^{(19)}(\pi-\vartheta), -\mathbf{Km}^{(19)}(\pi-\vartheta), \mathbf{Kn}^{(24)}(\pi-\vartheta), \mathbf{Km}^{(24)}(\pi-\vartheta)).\end{aligned} \tag{3.14}$$

Notice that the "slots" for rods 21 and 24 do not change in the actions (3.13), (3.14).

We choose the subgroup $\Sigma = \{\mathbf{I}, \mathbf{K}\}$ for the construction of the reduced problem (2.34). From (3.2) and (3.13), it then follows that $(\mathbf{Kr}^{(21)}(\pi - \vartheta), (q_o, -\mathbf{Kq})^{(21)}(\pi - \vartheta)) \equiv (\mathbf{r}^{(21)}(\vartheta), (q_o, \mathbf{q})^{(21)}(\vartheta))$. Evaluating this at mid-point P $(\vartheta = 3\pi/2 = -\pi/2)$ gives $(\mathbf{Kr}_P, (q_{oP}, -\mathbf{Kq}_P)) \equiv (\mathbf{r}_P, (q_{oP}, \mathbf{q}_P))$, implying

$$\begin{aligned} &\mathbf{i}_1 \cdot \mathbf{r}_P = 0,\ \mathbf{i}_\alpha \cdot \mathbf{q}_P = 0, \\ &\mathbf{i}_1 \cdot \mathbf{r}_Q = 0,\ \mathbf{i}_\alpha \cdot \mathbf{q}_Q = 0,\ \alpha = 2,3, \end{aligned} \tag{3.15}$$

the second set of which follows from the same arguments applied to rod 24 (evaluated at $\vartheta = \pi/2$). Similar conditions result at joint A. For instance, (3.2) and (3.13) via the reduced problem imply that $(\mathbf{Kr}^{(17)}(s), (q_o, -\mathbf{Kq})^{(17)}(s)) \equiv (\mathbf{r}^{(14)}(s), (q_o, \mathbf{q})^{(14)}(s))$. Evaluating this at point A $(s = 0)$ gives $(\mathbf{Kr}_A, (q_{oA}, -\mathbf{Kq}_A)) \equiv (\mathbf{r}_A, (q_{oA}, \mathbf{q}_A))$, implying

$$\mathbf{i}_1 \cdot \mathbf{r}_A = 0,\ \mathbf{i}_\alpha \cdot \mathbf{q}_A = 0,\ \alpha = 2,3. \tag{3.16}$$

The equilibrium conditions at P, Q and A follow from joint equilibrium and reflection symmetry. For example, at mid-point Q we generally have the equilibrium conditions $\mathbf{n}_Q + \mathbf{n}_{Q^+} = \mathbf{0}$ and $\mathbf{m}_Q + \mathbf{m}_{Q^+} = \mathbf{0}$, where $\mathbf{n}_Q, \mathbf{m}_Q$ refer to the force and moment resultant on the substructure, while $\mathbf{n}_{Q^+}, \mathbf{m}_{Q^+}$ refer to that on the right substructure. From $(3.11)_1$ and (3.14), the reduced problem implies $(\mathbf{Kn}^{(24)}(\pi - \vartheta), -\mathbf{Km}^{(24)}(\pi - \vartheta)) \equiv (\mathbf{n}^{(24)}(\vartheta), \mathbf{m}^{(24)}(\vartheta))$. Evaluating this at the mid-point Q $(\vartheta = \pi/2)$, we find $(\mathbf{Kn}_Q, -\mathbf{Km}_Q) \equiv (\mathbf{n}_{Q^+}, \mathbf{m}_{Q^+})$. Thus, the equilibrium conditions now read $\mathbf{n}_Q = -\mathbf{Kn}_Q$ and $\mathbf{m}_Q = \mathbf{Km}_Q$, implying

$$\begin{aligned} &\mathbf{i}_1 \cdot \mathbf{m}_Q = 0,\ \mathbf{i}_\alpha \cdot \mathbf{n}_Q = 0, \\ &\mathbf{i}_1 \cdot \mathbf{m}_P = 0,\ \mathbf{i}_\alpha \cdot \mathbf{n}_P = 0,\ \alpha = 2,3, \end{aligned} \tag{3.17}$$

the second of which follows from the same argument applied to rod 21. Balance of force and moments at joint A $(s = 0)$ are generally given by

$$\begin{aligned} &\mathbf{n}^{(13)}(0) + \mathbf{n}^{(14)}(0) + \mathbf{n}^{(15)}(0) + \mathbf{n}^{(16)}(0) + \mathbf{n}^{(17)}(0) + \mathbf{n}^{(18)}(0) = -\lambda \mathbf{i}_3, \\ &\mathbf{m}^{(13)}(0) + \mathbf{m}^{(14)}(0) + \mathbf{m}^{(15)}(0) + \mathbf{m}^{(16)}(0) + \mathbf{m}^{(17)}(0) + \mathbf{m}^{(18)}(0) = \mathbf{0}, \end{aligned} \tag{3.18}$$

respectively. For the reduced problem via $(3.11)_1$ and (3.14), we deduce $\mathbf{n}^{(16)}(0) = \mathbf{Kn}^{(15)}(0)$, $\mathbf{m}^{(16)}(0) = -\mathbf{Km}^{(15)}(0)$, $\mathbf{n}^{(17)}(0) = \mathbf{Kn}^{(14)}(0)$, $\mathbf{m}^{(17)}(0) = -\mathbf{Km}^{(14)}(0)$, $\mathbf{n}^{(18)}(0) = \mathbf{Kn}^{(13)}(0)$, and $\mathbf{m}^{(18)}(0) = -\mathbf{Km}^{(13)}(0)$. Hence, (3.18) reduces to

$$\begin{aligned} &[\mathbf{I} + \mathbf{K}](\mathbf{n}^{(13)}(0) + \mathbf{n}^{(14)}(0) + \mathbf{n}^{(15)}(0)) = -\lambda \mathbf{i}_3, \\ &[\mathbf{I} - \mathbf{K}](\mathbf{m}^{(13)}(0) + \mathbf{m}^{(14)}(0) + \mathbf{m}^{(15)}(0)) = \mathbf{0}. \end{aligned} \tag{3.19}$$

Note that $\text{rank}[\mathbf{I} + \mathbf{K}] = 2$ and $\text{rank}[\mathbf{I} - \mathbf{K}] = 1$, i.e., (3.19) yields 3 scalar equations. Conditions (3.15)-(3.17) and (3.19) provide the extra conditions needed for maintaining reflection symmetry on the substructure of Figure 12. In order to implement AUTO, all rods in the half-structure are rescaled according to (2.37). This includes the half-lengths of rods DP and FQ, which need to be rescaled according to (2.37) with $L =$ 3. As before, the field equations for each rod in the half-structure need to be written with respect to their respective local coordinate system, the details of which we omit.

The results for the half-structure are shown below in Figures 13 and 14. As expected, the simple $C_{3v}$-symmetric bifurcations at $\lambda =$ 1.15 and $\lambda =$ 1.22 (see Figures 9 and 10, respectively) are captured. The first new bifurcation detected occurs at $\lambda =$ 1.26, which is transcritical; we deduce that it is associated with irrep $\gamma^{(5)}$. A typical configuration along with the solution path is shown in Figure 13. Along this branch, the deformed structure appears to possess $C_{2v}$, i.e., rectangular symmetry. This will be confirmed shortly. The second new bifurcation occurring at $\lambda = 1.42$ is a subcritical pitchfork. Hence, it is associated with irrep $\gamma^{(6)}$. The computed branch at hand, depicted in Figure 14, is one of two non-equivalent solution branches occurring at that bifurcation point. This too will be confirmed shortly. A typical configuration along the computed solution path is also given in Figure 14.

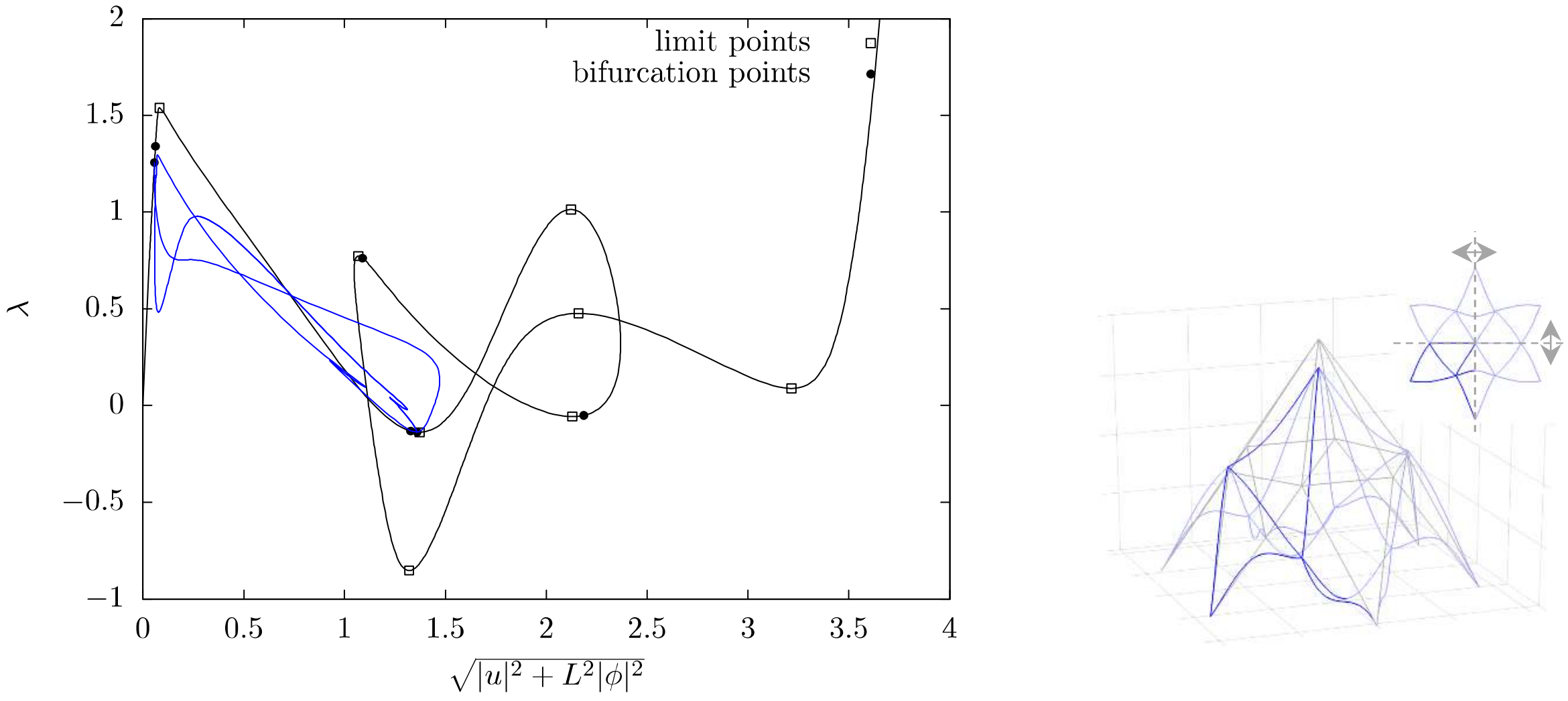

*Figure 13. Transcritical double bifurcation point at $\lambda = 1.26$ and typical configuration with $C_{2v}$ symmetry.*

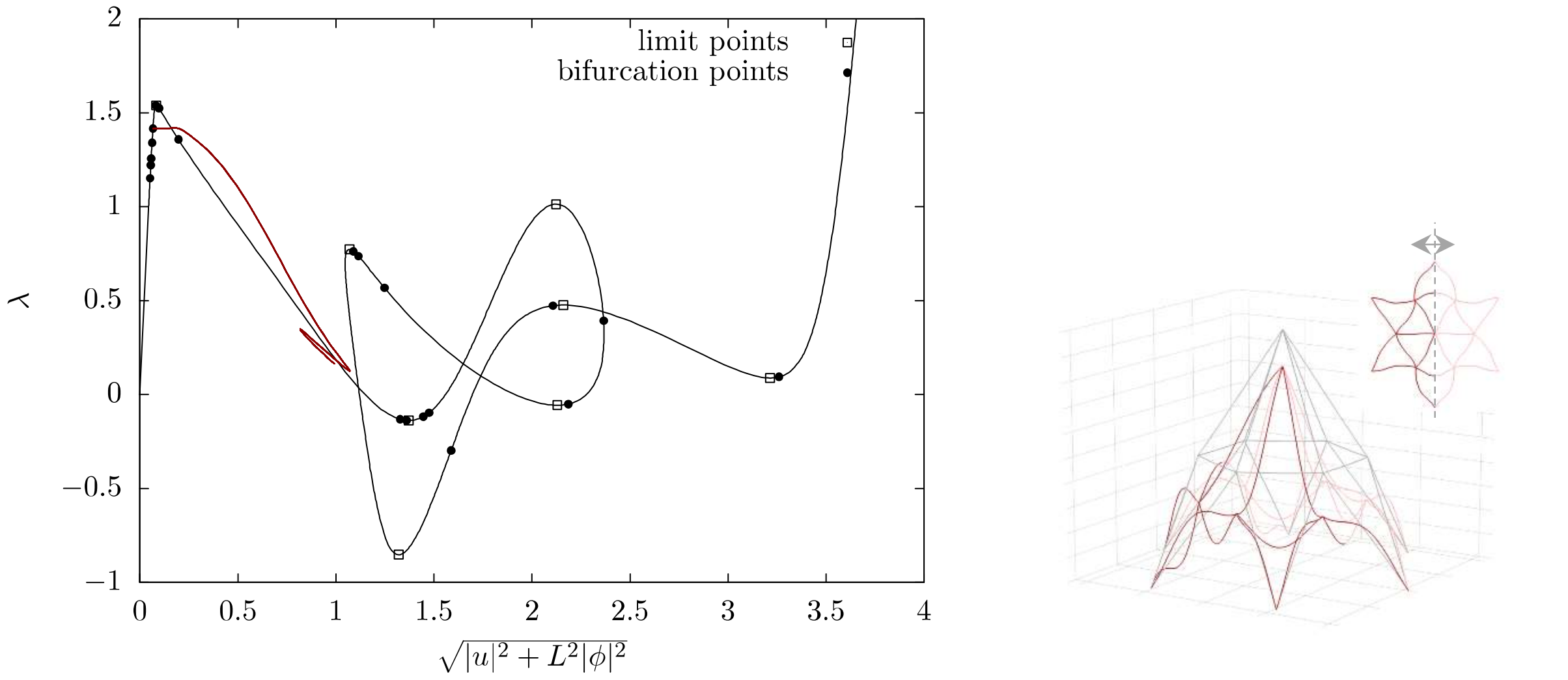

*Figure 14. Pitchfork double bifurcation point at $\lambda = 1.42$ and typical configuration with $Z_2$ symmetry.*

Next, we consider the reduced problem (2.34) corresponding to subgroup $\Sigma = \{\mathbf{I}, \mathbf{E}\}$; the associated substructure chosen here is depicted in Figure 15. As before, the boundary conditions at the supports and the interior joint compatibility and equilibrium conditions at joints F and G are the same used in the analysis of the complete structure. The boundary conditions at joints A, B and E arise from the reflection symmetry associated with the reduced problem. For instance, (3.2) and $(3.8)_2$ imply $(\mathbf{E}\mathbf{r}^{(22)}(s),(q_o,-\mathbf{E}\mathbf{q})^{(22)}(s)) \equiv (\mathbf{r}^{(23)}(s),(q_o,\mathbf{q})^{(23)}(s))$. Evaluating this at $s=0$, we deduce $(\mathbf{E}\mathbf{r}_B,(q_{oB},-\mathbf{E}\mathbf{q}_B)) \equiv (\mathbf{r}_B,(q_{oB},\mathbf{q}_B))$, and thus,

$$\mathbf{i}_2 \cdot \mathbf{r}_B = 0,\ \ \mathbf{i}_\alpha \cdot \mathbf{q}_B = 0,\ \ \alpha = 1\,\&\,3. \tag{3.20}$$

Similar considerations lead to the same type of conditions at joints A and E:

$$\begin{aligned} &\mathbf{i}_2 \cdot \mathbf{r}_A = 0,\ \ \mathbf{i}_\alpha \cdot \mathbf{q}_A = 0, \\ &\mathbf{i}_2 \cdot \mathbf{r}_E = 0,\ \ \mathbf{i}_\alpha \cdot \mathbf{q}_E = 0, \alpha = 1\,\&\,3. \end{aligned} \tag{3.21}$$

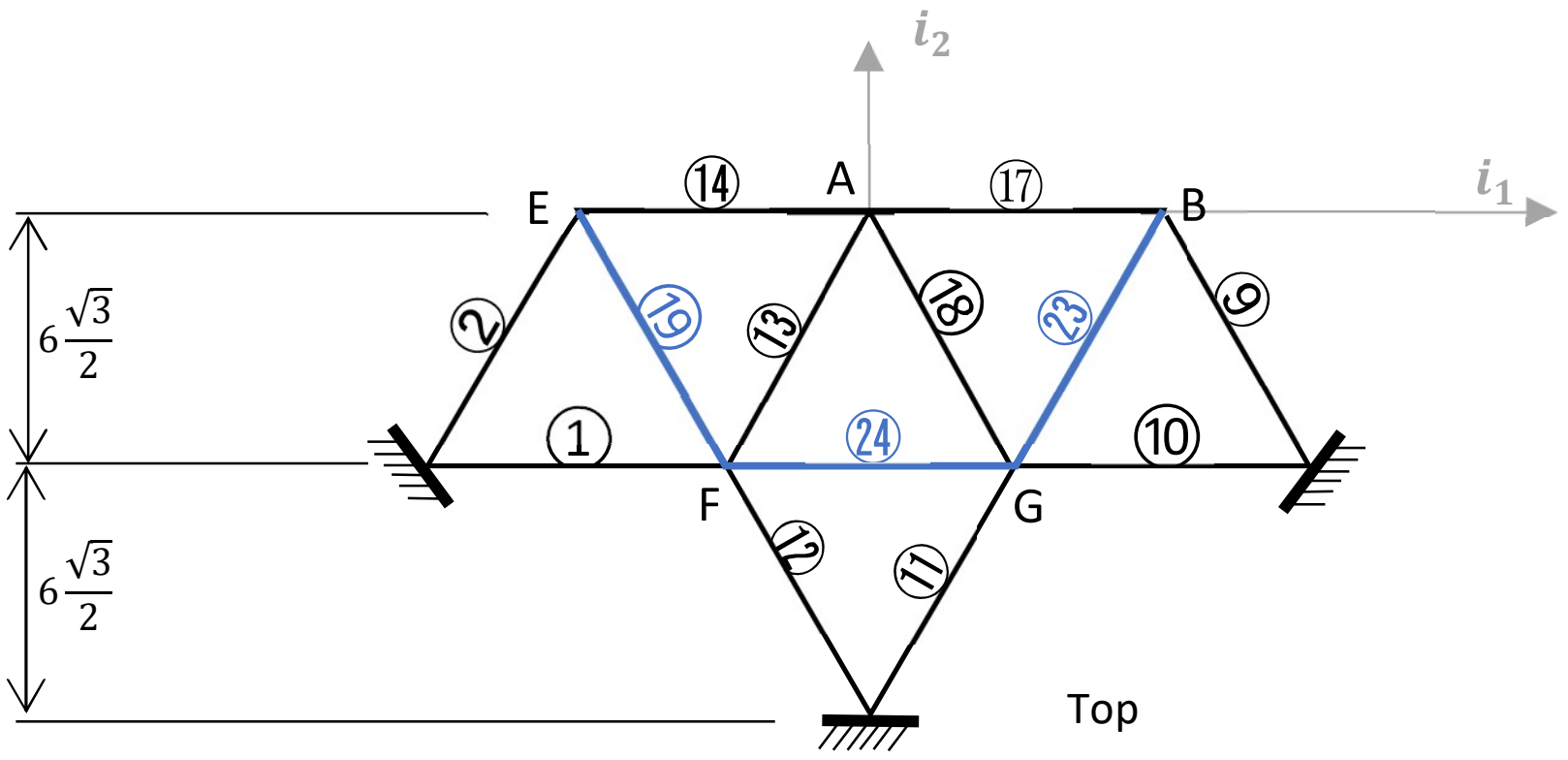

*Figure 15. Top view of the substructure associated with symmetry group* $\Sigma = \{\mathbf{I}, \mathbf{E}\}$.

Force and moment balance at joint E generally read

$$\begin{aligned} &\mathbf{n}^{(19)}(6) + \mathbf{n}^{(20)}(0) + \mathbf{n}^{(2)}(0) + \mathbf{n}^{(3)}(0) + \mathbf{n}^{(14)}(\sqrt{37}) = -\lambda \mathbf{i}_3, \\ &\mathbf{m}^{(19)}(6) + \mathbf{m}^{(20)}(0) + \mathbf{m}^{(2)}(0) + \mathbf{m}^{(3)}(0) + \mathbf{m}^{(14)}(\sqrt{37}) = \mathbf{0}, \end{aligned} \tag{3.22}$$

respectively, where we employ the local arc-length coordinates of each rod. For the reduced problem, conditions $(3.11)_{1,3}$ imply $\mathbf{n}^{(20)}(0) = \mathbf{E}\mathbf{n}^{(19)}(6)$, $\mathbf{m}^{(20)}(0) = -\mathbf{E}\mathbf{m}^{(19)}(6)$, $\mathbf{n}^{(3)}(0) = \mathbf{E}\mathbf{n}^{(2)}(0)$, $\mathbf{m}^{(3)}(0) = -\mathbf{E}\mathbf{m}^{(2)}(0)$, and $\mathbf{E}\mathbf{n}^{(14)}(6) = \mathbf{n}^{(14)}(6)$. Thus, (3.22) becomes

$$\begin{aligned} &[\mathbf{I}+\mathbf{E}]\{\mathbf{n}^{(19)}(6) + \mathbf{n}^{(2)}(0) + \mathbf{n}^{(14)}(\sqrt{37})/2\} = -\lambda \mathbf{i}_3, \\ &[\mathbf{I}-\mathbf{E}]\{\mathbf{m}^{(19)}(6) + \mathbf{m}^{(2)}(0) + \mathbf{m}^{(14)}(\sqrt{37})/2\} = \mathbf{0}. \end{aligned} \tag{3.23}$$

Beginning with the general equilibrium conditions (3.7), the reflection symmetry of the reduced problem at joint B implies

$$\begin{aligned} &[\mathbf{I}+\mathbf{E}]\{\mathbf{n}^{(23)}(0) + \mathbf{n}^{(9)}(0) + \mathbf{n}^{(17)}(\sqrt{37})/2\} = -\lambda \mathbf{i}_3, \\ &[\mathbf{I}-\mathbf{E}]\{\mathbf{m}^{(23)}(0) + \mathbf{m}^{(9)}(0) + \mathbf{m}^{(17)}(\sqrt{37})/2\} = \mathbf{0}. \end{aligned} \tag{3.24}$$

Finally, (3.18) and reflection symmetry lead to

$$[\mathbf{I}+\mathbf{E}]\{\mathbf{n}^{(13)}(0)+\mathbf{n}^{(18)}(0)+[\mathbf{n}^{(14)}(0)+\mathbf{n}^{(17)}(0)]/2\}=-\lambda\mathbf{i}_3,$$
$$[\mathbf{I}-\mathbf{E}]\{\mathbf{m}^{(13)}(0)+\mathbf{m}^{(18)}(0)+[\mathbf{m}^{(14)}(0)+\mathbf{m}^{(17)}(0)]/2\}=\mathbf{0}. \tag{3.25}$$

Note that rank$[\mathbf{I}+\mathbf{E}]$ = 2 and rank$[\mathbf{I}-\mathbf{E}]$ = 1. Thus, each of (3.23)-(3.25) represent 3 scalar equations.

The results for the half-structure of Figure 15 are as follows: As expected, the simple $C_{3v}$- symmetric solution branch, bifurcating at $\lambda =$ 1.31 and shown in Figure 11, is again found. Moreover, precisely the same transcritical bifurcation path depicted in Figure 13 is obtained. This confirms the $C_{2v}$- symmetry of the configurations along it. The underlying group-theoretic reasoning for this provided in Appendix B. Next, a new bifurcating pitchfork branch of solutions is detected at $\lambda = 1.42$, which is given in Figure 16 along with a typical configuration. Observe that this solution branch is distinct from that shown in Figure 14; e.g., each possesses a distinct reflection symmetry. This confirms that the double bifurcation point at $\lambda = 1.42$ is associated with irrep $\gamma^{(6)}$.

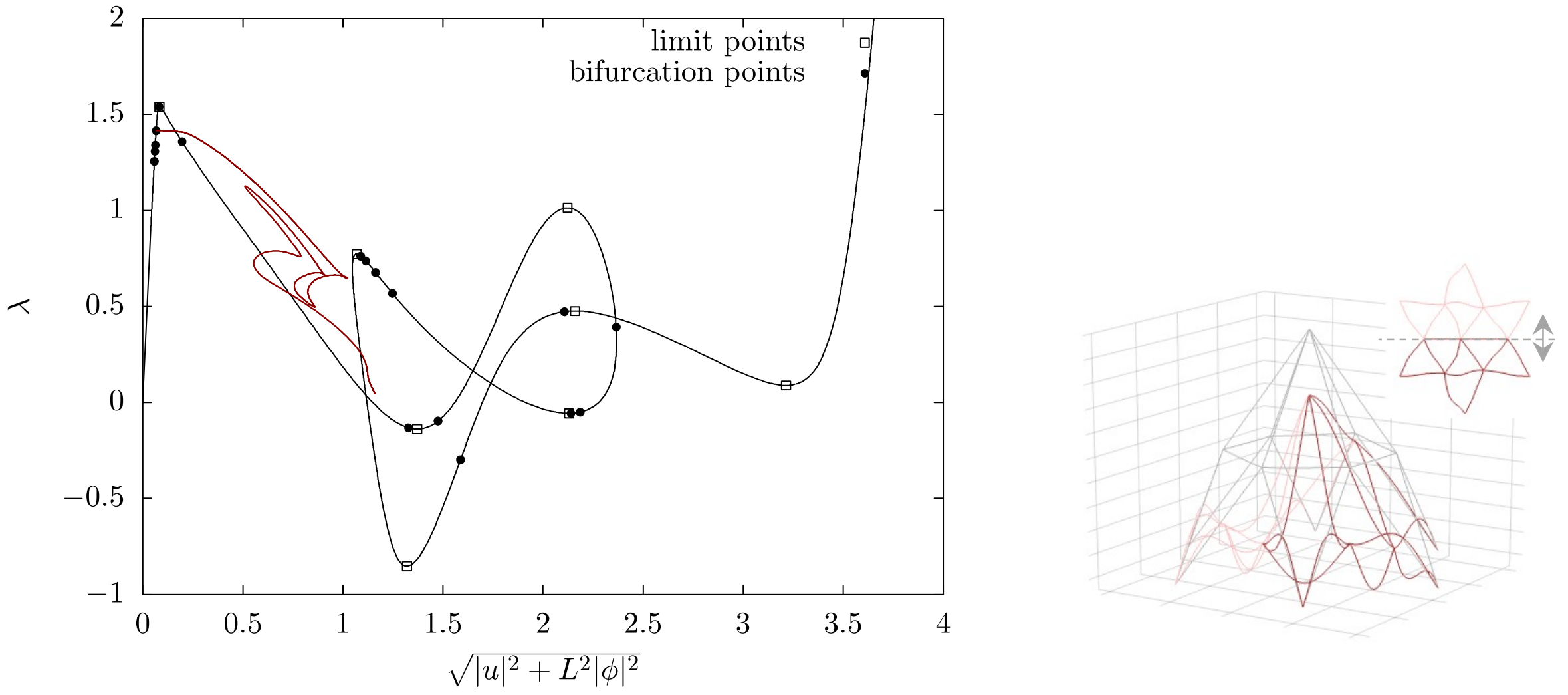

*Figure 16. Pitchfork double-bifurcation point at $\lambda = 1.42$ and typical configuration with $Z_2$ symmetry.*

Finally, we deduce some local stability information via the usual exchange-of-stability arguments [15]. *Stable* here means that the equilibrium is a local minimum of the total potential energy. The first ascending portion of the primary $C_{6v}$- symmetric path, between $\lambda = 0$ and $\lambda = 1.15$, is stable, after which stability is lost; the first subcritical $C_{3v}$- symmetric pitchfork emanating from $\lambda = 1.15$ is also unstable. However, the next subcritical $C_{3v}$- symmetric bifurcation at $\lambda = 1.22$ (precisely of the same type) regains stability as does the primary solution path between $\lambda = 1.22$ and $\lambda = 1.26$. Beyond that, the primary solution path and the remaining bifurcating solution paths - 3 branches at $\lambda = 1.26$, 1 branch at $\lambda = 1.31$, and 6 branches at $\lambda = 1.42$ - are all locally unstable.

## Concluding Remarks

We present a general approach to the bifurcation analysis of elastic frameworks with symmetry in this work. While group-theoretic methods for bifurcation problems with symmetry are well known, their actual implementation in the context of elastic frameworks is not straightforward. The main difficulty arises from the nonlinear configuration space, due to the presence of (cross-sectional) rotation fields. We avoid this via the single-rod formulation of [13], whereby the governing equations are embedded in a linear space. The field equations for a framework comprise the assembly of the rods equations, supplemented by compatibility and equilibrium conditions at the joints. We establish their equivariance via the invariance of the total potential energy under the group action. Moreover, the implementation of group-theoretic methods is natural within the linear-space context. All potential generic, symmetry-breaking bifurcations are a-priori predicted. We employ the numerical path-following code AUTO [5], which can detect and compute simple, one-dimensional bifurcations. For double bifurcation points, we construct symmetry-reduced problems implemented by appropriate substructures. The double bifurcations are rendered simple, and AUTO is again applicable. We mention that the subgroup-symmetry restrictions are imposed only at the exposed joints of the substructures used in Sections 2 and 3. In fact, the reduced problem (2.32), (2.35) impacts all exposed degrees of freedom – not just those at the exposed joints. Nonetheless, our practical approach employed here is demonstratively successful.

It is entirely possible to construct reduced problems via substructures for all simple bifurcations found as well – say for increased efficiency. For instance, this can be easily done for the tripod using a single rod. In the same way, each of the simple bifurcations for the hexagonal dome can be obtained via an appropriate substructure. But there is no real need for such, given that AUTO easily handles these for the complete structure.

Our formulation is not restricted to rigid (welded) joints. As indicated in (2.6) and the discussion immediately thereafter, other connection types are possible, e.g., ball-in-socket. Hinged connections (free rotation about one axis) and “mixed” connections at a joint are also possible. Moreover, our construction of reduced problems is not restricted to double bifurcation points; multiple bifurcation points arising from higher symmetry can be reduced in a similar manner. Finally, we point out that while our formulation is conveniently implemented via AUTO, it does not depend on it. Any two-point boundary solver – on a rod-by-rod basis, supplemented by joint equilibria and compatibility – can be employed as the basis for numerical continuation and for the detection/computation of simple bifurcations.

**Acknowledgements**

The work of TJH was supported in part by the National Science Foundation through grant DMS-2006586, which is gratefully acknowledged. CJC acknowledges support from Project ISITE FUTURE MAJOR, University Paris-Est Marne-la-Vallée.

## Appendix A

### Proof of Proposition 2.2:

First, we note that (2.28) is trivial for the symmetry-preserving loading term: From $(2.10)_1$, the first variation of the loading potential is simply $\lambda \mathbf{i}_3 \cdot \boldsymbol{\eta}_A \equiv \lambda \mathbf{G}\mathbf{i}_3 \cdot \boldsymbol{\eta}_A$ for all $\mathbf{G} \in C_{3v}$. So it's enough to show the validity of (2.28) for the internal potential energy $\mathcal{U}[\lambda, \mathbf{u}]$. Employing (2.9) and (2.29), we compute directional derivative of the invariance conditions (2.18) (ignoring the loading term):

$$\frac{d}{d\varepsilon}\tilde{\mathcal{U}}[\boldsymbol{T}_G\mathbf{u};\mathbf{h},\varepsilon)]|_{\varepsilon=0}=\frac{d}{d\varepsilon}\tilde{\mathcal{U}}[\mathbf{u};\mathbf{h},\varepsilon)]|_{\varepsilon=0}\Rightarrow$$
$$\left\langle\delta\mathcal{U}[\lambda,\boldsymbol{T}_G\mathbf{u}],\tilde{\boldsymbol{T}}_G\mathbf{h}\right\rangle=\left\langle\delta\mathcal{U}[\lambda,\mathbf{u}],\mathbf{h}\right\rangle, \tag{A.1}$$

for all $\mathbf{G}\in C_{3v}$ and for all admissible $\mathbf{h}$. The modified action on the space of admissible variations in $(A.1)_2$ arises as in (2.9) – from the conversion of skew-symmetric-matrix actions to cross products. It is straightforward, if tedious, to verify that the transformations (2.29) generate a faithful *representation* of $C_{3v}$, i.e., the mapping $\mathbf{G}\mapsto\tilde{\boldsymbol{T}}_G$ is one-to-one and

$$\tilde{\boldsymbol{T}}_G\tilde{\boldsymbol{T}}_H=\tilde{\boldsymbol{T}}_{GH},\ \tilde{\boldsymbol{T}}_G^{-1}=\tilde{\boldsymbol{T}}_{G^T},\ \tilde{\boldsymbol{T}}_I=\boldsymbol{I}\text{ (identity)}, \tag{A.2}$$

Moreover, we claim that

$$\tilde{\boldsymbol{T}}_G^*\equiv\tilde{\boldsymbol{T}}_G^{-1}\text{ on }C_{3v}, \tag{A.3}$$

where $\tilde{\boldsymbol{T}}_G^*$ denotes the adjoint operator of $\tilde{\boldsymbol{T}}_G$ with respect to the inner-product (2.30). To see this, first note that

$$\left\langle\tilde{\boldsymbol{T}}_G^*\tilde{\boldsymbol{T}}_G\boldsymbol{\sigma},\mathbf{h}\right\rangle=\left\langle\tilde{\boldsymbol{T}}_G\boldsymbol{\sigma},\tilde{\boldsymbol{T}}_G\mathbf{h}\right\rangle=\left\langle\boldsymbol{\sigma},\mathbf{h}\right\rangle\text{ on }C_{3v}, \tag{A.4}$$

for all smooth fields $\boldsymbol{\sigma},\mathbf{h}$, the second equality of which is readily verified via (2.29), (2.30). Thus, $\tilde{\boldsymbol{T}}_G^*$ is a left inverse. Now let $\boldsymbol{\sigma}=\tilde{\boldsymbol{T}}_G^*\boldsymbol{\mu}$ and $\mathbf{k}=\tilde{\boldsymbol{T}}_G\mathbf{h}$ in the second part of (A.4), leading to $\left\langle\tilde{\boldsymbol{T}}_G\tilde{\boldsymbol{T}}_G^*\boldsymbol{\mu},\mathbf{k}\right\rangle=\left\langle\tilde{\boldsymbol{T}}_G^*\boldsymbol{\mu},\mathbf{h}\right\rangle$ = $\left\langle\boldsymbol{\mu},\mathbf{k}\right\rangle$ on $C_{3v}$, for all $\boldsymbol{\mu},\mathbf{k}$, i.e., $\tilde{\boldsymbol{T}}_G^*$ is also a right inverse. Finally, we choose $\mathbf{h}=\tilde{\boldsymbol{T}}_G^*\mathbf{k}$ in $(A.1)_2$ and employ (A.3) to deduce

$$\begin{aligned}\left\langle\delta\mathcal{U}[\lambda,\boldsymbol{T}_G\mathbf{u}],\mathbf{k}\right\rangle&=\left\langle\delta\mathcal{U}[\lambda,\mathbf{u}],\tilde{\boldsymbol{T}}_G^*\mathbf{k}\right\rangle\\&=\left\langle\tilde{\boldsymbol{T}}_G\delta\mathcal{U}[\lambda,\mathbf{u}],\mathbf{k}\right\rangle,\end{aligned}$$

for all $\mathbf{G}\in C_{3v}$ and for all admissible $\mathbf{k}$. This completes the proof. □

**Appendix B**

Here we discuss the $C_{2v}$-symmetry of the transcritical bifurcation, depicted in Figure 13, associated with irrep $\gamma^{(5)}$. As pointed out above, $(2.36)_3$ and $(3.12)_5$ reveal that $\gamma^{(5)}$ is associated (faithfully) with $D_3$-symmetry on $\mathbb{R}^2$, but with each element appearing twice. Hence, the maximal isotropy subgroup of the fixed-point space $span\{\mathbf{i}_1\}$ is actually $\Sigma=\{\gamma_I^{(5)},\gamma_E^{(5)},\gamma_I^{(5)},\gamma_E^{(5)}\}$, which is a representation of $D_2:=\{\gamma_I^{(6)},\gamma_E^{(6)},\gamma_{R_\pi}^{(6)},\gamma_{ER_\pi}^{(6)}\}\subset D_6$. Accordingly, we can build a projection operator on configuration space with the symmetry of $\Sigma$ via

$$\boldsymbol{P}_\Sigma=\frac{1}{4}[\boldsymbol{T}_I+\boldsymbol{T}_E+\boldsymbol{T}_{R_\pi}+\boldsymbol{T}_{ER_\pi}],$$

where we use the $[\ ]_{11}$ entries of the matrices comprising $\Sigma$ for weights [17], all of which equal 1) in this case. The operators $\boldsymbol{T}_G$ are defined via the generators (3.8). Then $\boldsymbol{P}_\Sigma\boldsymbol{u}\in\mathcal{X}_o^\Sigma$, for any configuration $\boldsymbol{u}\in\mathcal{X}_o$, cf. (2.34), (3.2). We now observe that

$$\boldsymbol{T}_E\boldsymbol{P}_\Sigma = \boldsymbol{T}_E[\boldsymbol{T}_I + \boldsymbol{T}_E + \boldsymbol{T}_{R_\pi} + \boldsymbol{T}_{ER_\pi}]/4 = [\boldsymbol{T}_E + \boldsymbol{T}_I + \boldsymbol{T}_{ER_\pi} + \boldsymbol{T}_{R_\pi}]/4 = \boldsymbol{P}_\Sigma,$$
$$\boldsymbol{T}_{ER_\pi}\boldsymbol{P}_\Sigma = \boldsymbol{T}_{ER_\pi}[\boldsymbol{T}_I + \boldsymbol{T}_E + \boldsymbol{T}_{R_\pi} + \boldsymbol{T}_{ER_\pi}]/4 = [\boldsymbol{T}_{ER_\pi} + \boldsymbol{T}_{R_\pi} + \boldsymbol{T}_E + \boldsymbol{T}_I]/4 = \boldsymbol{P}_\Sigma. \tag{B.1}$$

Above we have used the fact that $\boldsymbol{T}_G\boldsymbol{T}_H = \boldsymbol{T}_{GH}$ for all $G, H \in C_{6v} \cong D_6$. Finally, for any $\boldsymbol{v} = \boldsymbol{P}_\Sigma \boldsymbol{u} \in \mathcal{X}_o^\Sigma$, (B.1) implies $\boldsymbol{T}_{R_\pi}\boldsymbol{v} = \boldsymbol{v}$, $\boldsymbol{T}_E\boldsymbol{v} = \boldsymbol{v}$, and $\boldsymbol{T}_{ER_\pi}\boldsymbol{v} = \boldsymbol{v}$, i.e., the configuration possesses $C_{2v}$ symmetry.